\documentclass[twocolumn,showpacs,preprintnumbers,amsmath,amssymb,pra]{revtex4-1}
\usepackage{graphicx}

\usepackage{color}
\begin{document}

\preprint{APS}

\title{Theoretical description of field-assisted post-collision interaction in Auger decay of atoms}

\author{S.~Bauch}
 \email{bauch@theo-physik.uni-kiel.de}
\author{M.~Bonitz}
\affiliation{%
Institut f\"ur Theoretische Physik und Astrophysik\\
Christian-Albrechts-Universit\"at zu Kiel, D-24098 Kiel, Germany
}%

\date{\today}

\begin{abstract}
In a recent publication [B. Sch\"utte, S. Bauch et al., accepted for publication in Phys. Rev. Lett. 2012] it was demonstrated both,
experimentally and theoretically, that Auger electrons are subject to an energetic chirp if a 
three-body-interaction dubbed post-collision interaction (PCI) is involved.
Here, we extend previous theoretical work and give a detailed analysis of field-assisted PCI
based on numerical solutions of the time-dependent Schr\"odinger equation, extensive Monte-Carlo
averaged molecular dynamics simulations and analytical theory.
The dependence on various streaking and excitation conditions is investigated, and we discuss, 
how these findings may help to improve XUV pulse characterization as well as understanding
of ultrafast atomic processes.
\end{abstract}



\maketitle

\section{Introduction}

The progress in the creation of phase stabilized laser systems and the generation of short and ultrashort pulses in the 
ultraviolet (UV) and extreme ultraviolet (XUV)
regime \cite{brabec_intense_2000,agostini_physics_2004} allows nowadays for the observation of electronic processes
on the femtosecond and even sub-femtosecond timescale in a time-resolved fashion \cite{scrinzi_attosecond_2006,krausz_attosecond_2009}.
Fundamental investigations include the mapping of the oscillating electrical field of a laser \cite{goulielmakis_direct_2004},
electron tunneling in strong fields \cite{uiberacker_attosecond_2007} or the direct observation of Auger decay in 
the time domain \cite{drescher_time-resolved_2002}. Recently, processes down to a duration of several tens of attoseconds have been
 demonstrated to be resolvable \cite{schultze_delay_2010}. 	

The major tool for the observation of fast processes since early days in physics is the \emph{streak-camera}.
Following its first mechanical realization by Wheatstone 1834 \cite{wheatstone1834} with $\mu$s resolution,
nowadays the sub-picosecond regime can be accessed with classical optoelectronic setups \cite{bradley1971,feng_x-ray_2007}.
To overcome the mechanical and electronic barriers for switching times, the answer was found in electrodynamics,
leading to a setup called light-driven streak camera \cite{kienberger_atomic_2004,itatani_attosecond_2002}:
Here, the temporal deflection of electrons is realized by the time varying vector potential of a laser field and 
the triggering of the process is done by ultrashort ionization through attosecond XUV pulses 
in pump-probe setups. A possibility to reach the zeptosecond regime in ultrahigh fields has recently
been proposed \cite{ipp_streaking_2011} theoretically.

An important application of the light-field driven streak camera is the characterization of (X)UV pump pulses 
in the femtosecond \cite{fruhling_single-shot_2009} and sub-fs regime \cite{itatani_attosecond_2002}.
By means of photoionization of rare-gas target atoms the XUV pulse properties, such as duration, substructure and chirp, 
are imprinted on a photoelectron distribution.
A time-varying streaking field deflects these electrons and maps the temporal
 properties to a measurable energy spectrum.
Therefore, this procedure strongly relies on the precise 
knowledge of the photon-to-electron conversion and, with that, a method to extract 
the temporal pulse properties from the
streaked kinetic energy spectra of the electrons.
While for solely photoelectrons this mapping is agreed to be understood for atoms
\cite{itatani_attosecond_2002,fruhling_single-shot_2009} and atoms on surfaces \cite{krasovskii_spectral_2007,krasovskii_towards_2009},
the situation strongly differs if Auger decay is involved.

The radiationless decay of resonances, first described by Lise Meitner 1922 \cite{meitner_ueber_1922} and Pierre Auger 1925 \cite{auger_sur_1925}, 
is a fundamental correlation-driven many-body effect in quantum mechanics, covering atoms, solids,
quantum dots and molecules. The photoexcited inner-shell-hole is subject to spontaneous decay, transferring its
energy to an outer-shell electron, that leaves the ion with its excess energy. 
The result is a doubly charged ion and two correlated electrons in the continuum.
Due to the spontaneous character of the hole decay, the Auger electron cannot carry information about the 
pump pulse.
Nevertheless, in \cite{schuette_evidence_2011} it was demonstrated, utilizing the THz streak camera setup,
that the energy of the Auger electron depends on its release time, i.e. it carries an energetic chirp,
if the Auger electron is faster than the preceding photoelectron. 
This could be verified using two independent experiments involving XUV photons from (i) the free electron Laser FLASH at DESY (Hamburg)
and (ii) a higher-harmonics generation (HHG) source \cite{schuette_electron_2011}. 
While for (i) the ionizing pulse has a complicated structure, both
in time and energy, for (ii) chirp-free pulses with rather well-defined properties are expected.
Nevertheless, it was established experimentally that for both, (i) and (ii), the Auger electron's chirp 
is present and has qualitatively the same properties.

The authors of \cite{schuette_evidence_2011} identified post-collision interaction (PCI) 
as the responsible mechanism for the observed chirp, utilizing extensive molecular dynamics (MD) simulations
as well as an analytically solvable model.
PCI is a process, where a fast Auger electron can catch up with the slower photoelectron, which leads to a drastic change
of the screening of the ion's charge. This manifests itself in an energy exchange: the photoelectron loses
energy (increased binding), whereas the Auger electron is correspondingly accelerated (the binding potential becomes shielded).
Obviously, the net amount of transferred energy depends on the distance from the ion, the closer the
overtaking happens the stronger the effect.
Although widely discussed in the literature \cite{niehaus_analysis_1977,
ogurtsov_auger_1983, russek_post-collision_1986,kuchiev_resonant_1986,kuchiev_post-collision_1989,aberg_unified_1992}
the consequences of PCI for the temporal energy distribution remained unexplored.
In this paper, we extend the theory presented in \cite{schuette_evidence_2011} and give a detailed description of 
field-assisted PCI (FA-PCI) using quantum and classical simulations as well as analytical theory.

The paper is organized as follows: 
In the first part, Sec.~\ref{sec:tdse}, we demonstrate the presence of a chirp on the Auger electron energy by solving the time-dependent
Schr\"odinger equation (TDSE) for model systems and support the idea of PCI being the responsible mechanism.
To overcome the model character necessary for quantum calculations, we develop in Sec.~\ref{sec:semi-classics} a classical
simulation technique based on Monte-Carlo (MC) averaged molecular dynamics (MD). In Sec.~\ref{sec:analytic}, extending
the model of \cite{schuette_evidence_2011},  we present an analytical theory for the Auger line shape in the presence 
of a slowly varying streaking field including
PCI effects and compare to TDSE as well as MC-MD simulations.
In Sec.~\ref{sec:params}, we investigate the influence of various pulse parameters and show, how the measurement of the 
PCI-induced chirp may help to
improve the pulse characterization capabilities of light-field-driven streak camera setups.
The paper closes with a comment on recent experiments and an outlook on future investigations in Sec.~\ref{sec:conclusions}.

\section{Quantum theory of laser-assisted Auger decay}
\label{sec:tdse}
Starting point for the description of laser-assisted Auger decay (LAAD) on a quantum mechanical level
 is the time-dependent Schr\"odinger equation (TDSE).
However, full quantum calculations of autoionization involve two or more electrons, which limits
them to model studies, see e.g. \cite{haan_numerical_1994}, or Helium \cite{hu_time-dependent_2005} on very
short time scales. 
To overcome this ``brute-force`` approach we 
use a generalization of Fano's theory \cite{fano_effects_1961} to the time-dependent case 
developed in 
\cite{kazansky_nonstationary_2005,kazansky_triple_2006,kazansky_time-dependent_2009} and references therein.
 The notations follow \cite{kazansky_time-dependent_2009}, an analogous derivation based on quantum field theory
can be found in \cite{buth_theory_2009}. A similar theoretical approach to time-resolved Fano resonances
is developed in \cite{wickenhauser_theoretical_2004,wickenhauser_time_2005}.

The time evolution of the outgoing photoelectron after excitation with an XUV pulse is governed by
a set of coupled TDSEs 
(throughout atomic units, $m_e=|e|=\hbar=4\pi\epsilon_0 \equiv1$, are used):
\begin{widetext}
\begin {eqnarray}
 i \frac{\partial}{\partial t} \phi_d(\boldsymbol{r},t)  &= & \left ( \hat{H}_1(\boldsymbol{r}) - i\frac{\Gamma_A}{2} - z E_L(t) \right ) \phi_d(\boldsymbol{r},t) - z E_X(t) \phi_0(\boldsymbol{r}) e^{-i \epsilon_0 t}
 \label{eq:systdse-1} \; , \\
 i \frac{\partial}{\partial t} \phi_{\epsilon} (\boldsymbol{r},t) &= &\left (\hat{H}_2(\boldsymbol{r}) - E_A + \frac{1}{2} [\boldsymbol{k}_A - \boldsymbol{A}_L(t)]^2 - z E_L(t) \right) \phi_{\epsilon}(\boldsymbol{r},t) + V \phi_d(\boldsymbol{r},t) \;.
\label{eq:systdse-2}
\end {eqnarray}
\end{widetext}
Eq.~\eqref{eq:systdse-1} describes the photoelectron excited from the initial orbital $\phi_0$ with 
energy $\epsilon_0$ by a laser pulse $E_X(t)$. It moves in a potential of a singly charged ion, included in $\hat{H}_1(\boldsymbol{r})$, and the streaking field
$E_L(t)$. In other words, $\phi_d(t)$ describes the photoelectron before decay of the resonance with decay constant $\Gamma_A$.
After Auger decay  with excess energy $E_A$, being the energy difference between the outer shell electron and the core hole,
 the photoelectron's movement in the potential of a doubly charged ion, 
contained in $\hat{H}_2$, is governed by Eq.~\eqref{eq:systdse-2} coupled to Eq.~\eqref{eq:systdse-1} via the Auger
 decay matrix element $V$, which is assumed to be constant in energy and space~\cite{kazansky_time-dependent_2009}.
 By setting $\hat{H}_1=\hat{H}_2$, post-collision effects due to changed screening of the ion's charge
can be artificially excluded from the calculations.

In fact, Eq.~\eqref{eq:systdse-2} represents a set of equations for all possible energies of the Auger electron
$\epsilon=\boldsymbol{k}_A^2/2$. The vector potential $\boldsymbol{A}_L(t)$ associated with the electrical field $\boldsymbol{E}_L(t)$,
\begin{equation}
 \boldsymbol{A}_L(t)=-\int_{-\infty}^{t} \; \textup{d}\tau E_L(\tau) \;,
\end{equation}
is chosen to vanish for long times.

Both laser pulses are  linearly polarized in $z$ direction with Gaussian envelopes 
and coupled to Eqs.~\eqref{eq:systdse-1} and~\eqref{eq:systdse-2} in dipole approximation.
 The streaking pulse with duration $\tau_L$, phase shift $\varphi_L$ and frequency $\omega_L$ is centered at zero,
\begin{equation}
 \boldsymbol{E}_L(t)= \hat{e}_z E^0_L \exp\left(-\frac{t^2}{2 \tilde{\tau}_L^2} \right) \cos \left[ \omega_Lt+\varphi_L\right] \;.
 \label{eq:EL}
\end{equation}
The XUV pulse is delayed by $t_X$ with photon energy $\omega_X$ and duration $\tau_X$,
\begin{equation}
 \boldsymbol{E}_X(t)=\hat{e}_z E^0_X \exp\left(-\frac{(t-t_X)^2}{2 \tilde{\tau}_X^2} \right ) \cos \left [\omega_X (t-t_X) \right] \;.
 \label{eq:EX}
\end{equation}
Note: throughout this paper, all pulse durations are given as
full width at half maximum (FWHM) and will be denoted by $\tau_{X,L}=2\sqrt{2 \ln 2}\tilde{\tau}_{X,L}$.
The model, Eqs.~\eqref{eq:systdse-1} and \eqref{eq:systdse-2}, has been successfully applied to the recapture of photoelectrons due to PCI 
\cite{hergenhahn_population_2006,hergenhahn_study_2005} 
and to (angle-resolved) sideband structures in LAAD \cite{kazansky_sideband_2009,kazansky_gross_2010}, which 
appear if the duration of the pump pulse is comparable to or longer than the period of the streaking field.
In this work, we use  $\tau_X\ll 1/\omega_L$ required for streak cameras.

\subsection{Simplifications}

Up to now, the above-mentioned previous works considered  short pulses in the (sub-) fs regime involving infrared (IR)
streaking pulses. The characterization of pump pulses longer  than $20\,$fs, as they are produced e.g. by free electron lasers,
 requires deflecting fields based  on
THz radiation \cite{schuette_evidence_2011,fruhling_single-shot_2009} and, therefore, 
requires the propagation of Eqs.~\eqref{eq:systdse-1} and \eqref{eq:systdse-2} over a duration of several picoseconds. 
In order to describe the involved processes on a time-dependent quantum mechanical level
 drastic simplifications are needed to keep the computational costs manageable.

As a first step, we restrict our investigations to a one-dimensional (1D) version, i.e. consider
wave functions of the form $\phi_d(x,t)$ and $\phi_{\epsilon}(x,t)$, neglecting any angular momenta 
and distributions.
This leads to the model Hamiltonians $\hat{H}_1(x)$ and $\hat{H}_2(x)$ which are chosen to account for the correct 
asymptotics of the binding potentials of the remaining ion,
\begin{equation}
  \hat{H}_i(x)=-\frac{1}{2} \frac{\partial^2}{\partial x^2} + \frac{Z_i}{\sqrt{x^2+\kappa^2}} \;,
  \label{eq:1d-hamiltonian} 
\end{equation}
with $Z_1=-1$ and $Z_2=-2$. The Coulomb singularity appearing in 1D systems
has been regularized in a standard procedure by $\kappa$, e.g. \cite{haan_numerical_1994,su_model_1991,bauch_electronic_2010}, 
assuring a finite binding potential at the position of the ion.

Still, to keep track of the photoelectrons traveling with $25$ to $80\,$eV in the continuum, enormous computational
grids are needed. To overcome this point, we introduce a scaling procedure of all relevant temporal quantities
by a factor $\gamma$, which maps the (not-manageable) physical system to a smaller-sized analog, which can
be tackled by the quantum simulations:
\begin{equation}
 \Gamma_A^*=  \gamma \Gamma_A, \; \omega_L^* = \gamma\omega_L, \; \tau_X^*=\frac{\tau_X}{\gamma},\; \tau_L^*=\frac{\tau_L}{\gamma} \; .
 \label{eq:scaling}
\end{equation}
In order to keep the relevant streaking conditions comparable, the intensity of the streaking field is chosen such 
that the ponderomotive potential 
$U_p=E_0^2/4 \omega_L^2$ of the streaking field is kept constant when $\gamma$ is varied.
The influence of this scaling procedure is discussed in detail below.

\subsection{Solution of Eqs.~\eqref{eq:systdse-1} and \eqref{eq:systdse-2}}
\begin{figure}
 \includegraphics[width=8.5cm]{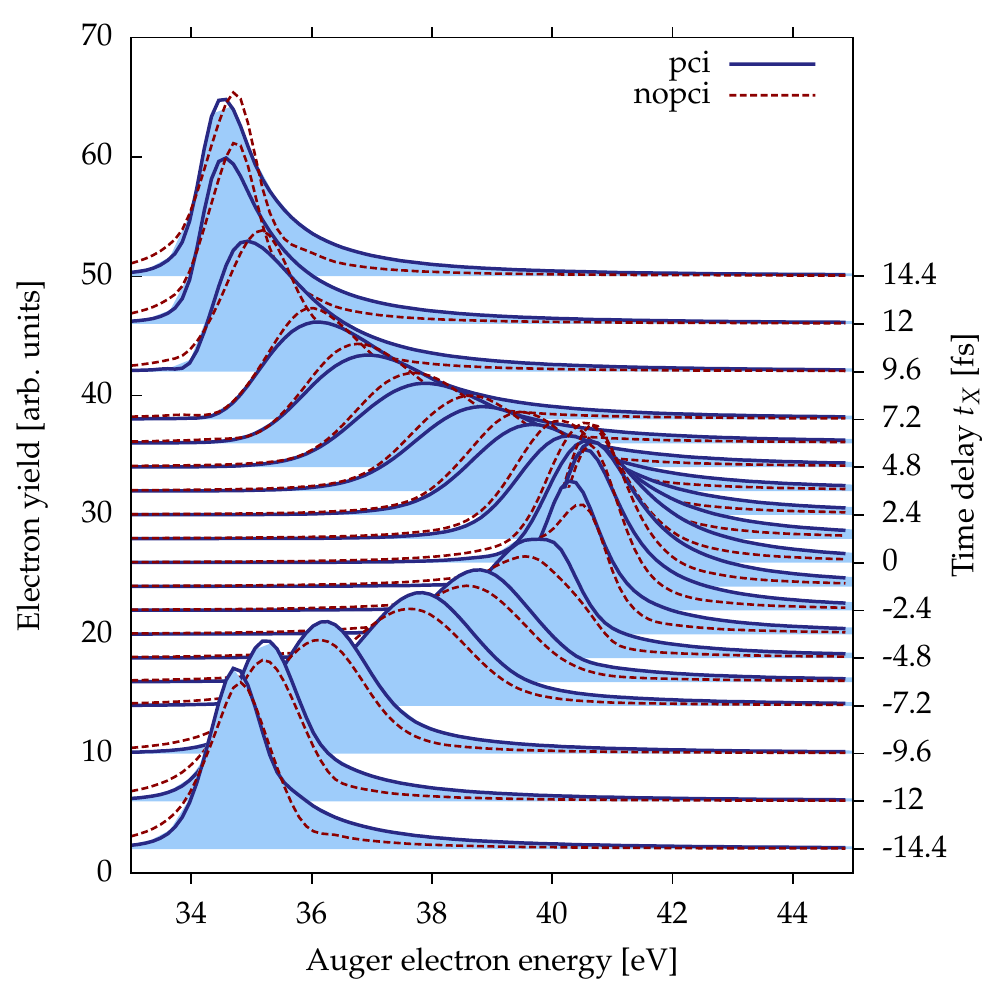}
 \caption{(color online) Auger line shapes obtained by solving Eqs.~\eqref{eq:systdse-1} and~\eqref{eq:systdse-2} for a
series of time  delays $t_X$. Shown is the case including PCI (blue solid lines, blue area) and the
case neglecting PCI (red dashed lines),  i.e. $\hat{H}_2=\hat{H}_1$ in Eqs.~(\ref{eq:systdse-1}, \ref{eq:systdse-2}) for
a $16.9\,$fs single-cycle ($\varphi_L=\pi/2$) streaking pulse with a frequency of $\omega_L=33\,$THz and  $U_p=98.8\,$meV.
The $2.83\;$fs XUV pulse has a photon energy of $91\;$eV.
The Auger decay constant was set to $\Gamma_A=950\,$meV at an Auger energy of $E_A=34.27\,$eV thus resembling
THz-streaking of the Xe \emph{NOO} transition scaled by a factor of $\gamma=10$ (see text for details).
Note: the lines neglecting PCI have been shifted towards higher energy by $0.5\,$eV for better comparison.}
 \label{fig:tdse-lines}
\end{figure}

We solve the 1D analogs of Eqs.~\eqref{eq:systdse-1} and~\eqref{eq:systdse-2} employing a finite-element discrete variable
representation (FE-DVR) \cite{rescigno_numerical_2000, schneider_parallel_2006} 
 and an independent finite-difference based method on large spatial grids allowing for the propagation of several tens of fs 
without reflections at the grid edges. All considered observables have been carefully checked
for convergence with respect to the numerical discretizations.

Throughout this paper, two transitions motivated by the experiment are considered,
 the \emph{NOO} transition in Xenon and the \emph{MNN} transition in Krypton \cite{schuette_evidence_2011}.
Let us start with Xenon.
For that, the eigenstate of $\hat{H}_1$  ($\kappa=0.1935$) 
with a ground state energy of $E_p=-66\,$eV is used for XUV excitation with a photon energy of $w_X=91\,$eV
which corresponds to a kinetic energy of $w_X+E_p=25\;$eV for the photoelectron.
The Auger electron energy is chosen to match $E_A=34\,$eV, being faster than the photoelectron wave packet and thus giving rise to PCI effects.
An example of the resulting Auger electron line shapes for a full scan of time delays $t_X$ is shown in Fig.~\ref{fig:tdse-lines}
for a streaking with $33\,$THz and $U_p=98.8\,$meV and a pump pulse duration of $2.83\,$fs, thus scaled by a factor of $\gamma=10$
in comparison with the THz streak camera in \cite{fruhling_single-shot_2009}. 
Analogously, the atomic parameters, matching the Xe \emph{NOO}
transition, are scaled  by the same factor according to Eq.~\eqref{eq:scaling}.

Each individual Auger line is shown for two cases: including PCI (blue solid lines) and neglecting PCI (red dashed lines).
For both cases, the typical streaking picture of the time-dependent momentum transfer arises, with a
general shift of the PCI result towards higher energies ($0.5\;$eV), which is compensated in Fig.~\ref{fig:tdse-lines} for better
visibility. 
Careful inspection of the line shapes reveals, that for falling slope of $A(t)$ ($t_X<0$) the lines are higher and
 of smaller width than for the case of rising slope of $A(t)$ ($t_X>0$), cf. pci vs. nopci curves. Note, that the
energy shift is proportional to $-A(t)$.
The lines corresponding to the case without PCI have the same height and width for positive and negative time delays.

This observation already indicates a chirp in Auger electron emission, i.e.~a
time-dependent variation of the energy of the Auger electron manifesting itself in an asymmetry with respect to the direction of the slope of $A$.
In the following, this result will be investigated in detail and the underlying physical mechanism will be identified.

\subsection{Analysis of the TDSE results}
\label{ssec:tdse-results}

\begin{figure}
 \includegraphics[width=0.49\textwidth]{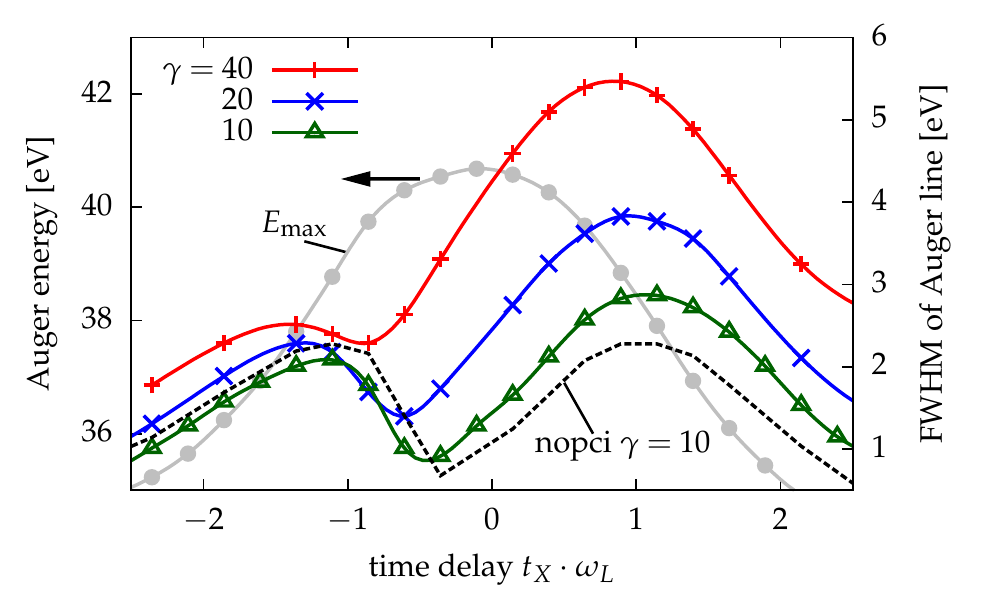}
 \caption{(color online) FWHM of Auger lines (right axis) obtained by solving Eqs.~\eqref{eq:systdse-1} and \eqref{eq:systdse-2} for different scaling factors 
$\gamma$, cf. Eq.~\eqref{eq:scaling}, of the Xe \emph{NOO} transition in a $3.3\,$THz streaking field
keeping $U_p=98.8\,$meV constant.
 The natural line width $\Gamma_A$ has been subtracted for each set of parameters for better 
comparison. The maximum of the line (proportional to $-A(t)$, gray line with filled circles labeled by $E_{\textup{max}}$, left axis) and 
the case neglecting PCI (black dashed line) are shown for $\gamma=10$. 
For $\gamma=10$, parameters are the same as in Fig.~\ref{fig:tdse-lines}.}
 \label{fig:scaling}
\end{figure}
Let us first discuss the influence of the scaling procedure \eqref{eq:scaling}, shown in Fig.~\ref{fig:scaling}.
We point out, that each value of $\gamma$ corresponds to a certain physical system, but our aim
is to describe experiments based on the Xe \emph{NOO} transition.
The width displayed in Fig.~\ref{fig:scaling} is extracted from line shape data by interpolation utilizing cubic splines and 
subsequent finding of the maximum and the corresponding FWHM.
For better comparison, the $x$-axis is shifted  by the Auger decay time  $\Gamma_A^{-1}$ for each data set 
and the width was modified
by $\sqrt{\sigma^2-\Gamma_A^2}$ to account for the different natural line width in each set of parameters.
The first observation is a strong asymmetry in the FWHM for all values 
of $\gamma$ with respect to the slope of $A(t)$. Note: the displayed curve labeled $E_{\textup{max}}$
shows the energy corresponding to the maximum of the Auger line, which is proportional to $-A(t)$.
Approaching the physical system of Xe \emph{NOO} ($\gamma=1$), the asymmetry gets smaller, but is still present
for the smallest considered value of $\gamma$.
For comparison, also the case neglecting PCI  is shown for $\gamma=10$, where no
 such asymmetry is observed and the typical chirp-free streaking
behavior \cite{itatani_attosecond_2002} is retrieved: 
largest width (and corresponding time resolution of the streak camera) occurs at maximum slope
of $A(t)$.   Note: for single-cycle pulses used here, this does not coincide with zero transitions 
of $A(t)$ (electrical field maxima). 
At the maximum of $|A(t)|$,  as expected,  a pronounced minimum can be observed.
\begin{figure}
   \includegraphics[width=8.5cm]{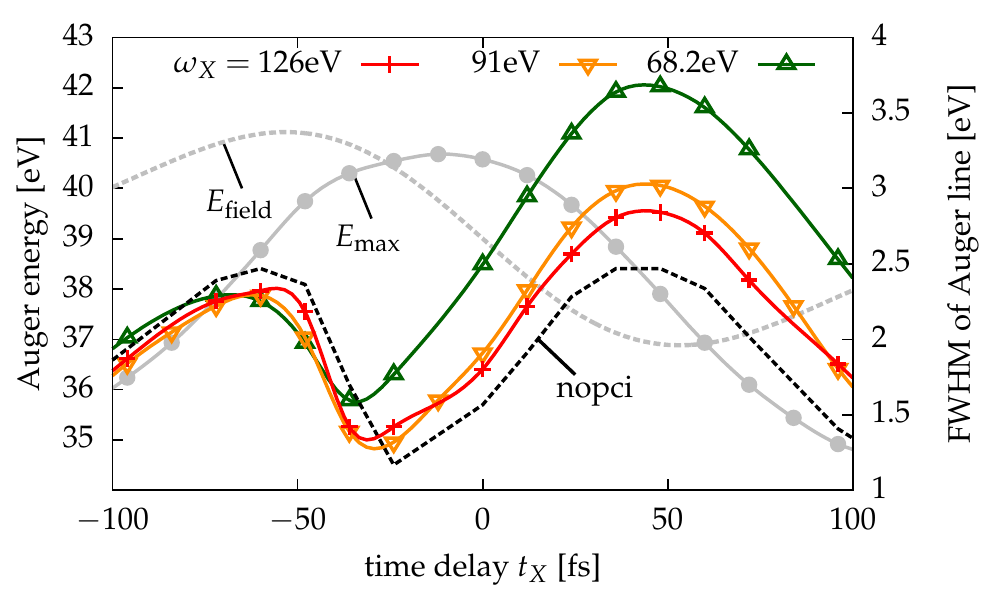}
   \caption{(color online) FWHM of Auger lines (right axis) for different photon energies $\omega_X$ of the XUV pulse of $28\,$fs
 duration. All other parameters are the same as in Fig.~\ref{fig:scaling} for the case $\gamma=10$.
The energy of the maximum of the Auger line and the case neglecting PCI 
(black dashed line) are given for $\omega_X=91\,$eV. The time dependence of the 
electrical field $E(t)$ is sketched by the gray dotted line.}
   \label{fig:tdse-pevel}
\end{figure}

Since the PCI effect originates from the changed screening of the remaining ions' charge during overtaking of
the photoelectron by the Auger electron, it strongly depends on the velocity of the photoelectron, cf. Sec.~\ref{ssec:lineshapes-pci}.
 Therefore, the observed asymmetry
should be more pronounced for slow photoelectrons, where the overtaking happens in close vicinity to the ion \cite{schuette_evidence_2011}, and should vanish
for fast ones, where the Auger electron cannot catch up with the photoelectron.
Fig.~\ref{fig:tdse-pevel} shows the FWHM of the Auger line of the Xe \emph{NOO} transition ($\gamma=10$) for a set of 
photon energies $\omega_X$.
As is clearly seen, the strongest asymmetry is observed for slow photoelectrons (green curve with triangles)
whereas the increase of the photoelectron's energy leads to a decrease of the observed asymmetry in the FWHM and approaches the
case neglecting PCI (black dashed line), thus supporting the idea that PCI is responsible for the energetic chirp in Auger emission.
We note that, although for $\omega_X=126\,$eV rather fast photoelectrons ($47\,$eV in comparison to $35\,$eV Auger electron energy) are emitted (red curve), 
still an asymmetry is observed. 
This originates from a rather broad distribution of the photoelectron energy.

\begin{figure}
  \includegraphics[width=8.5cm]{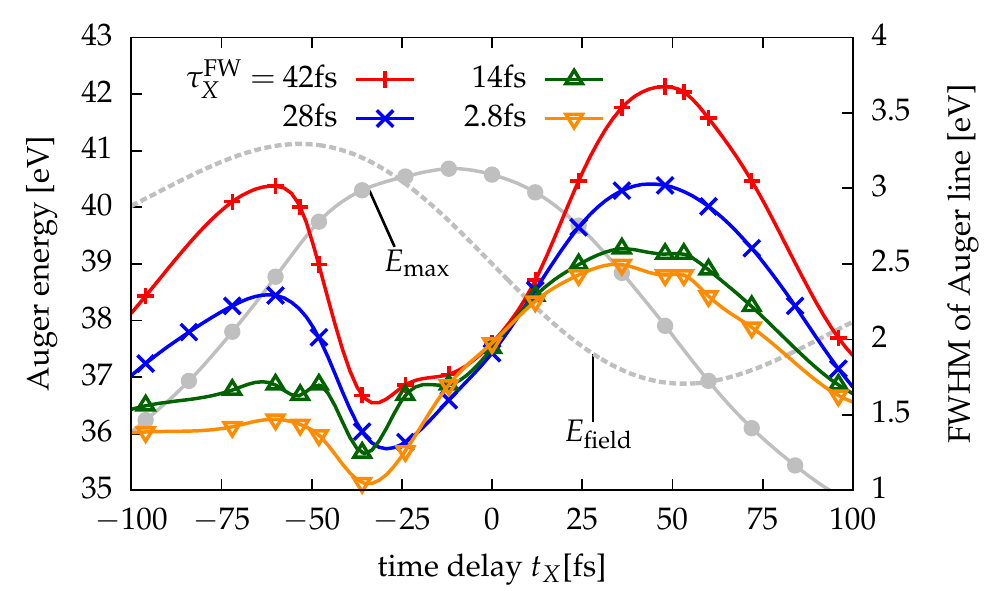}
  \caption{(color online) FWHM of Auger line (right axis) for different XUV pulse durations $\tau_X^{\textup{FW}}$.
The maximum of the Auger line (proportional to $-A(t)$, filled circles) is shown for $28\,$fs.
The time dependence of the electrical field $E(t)$ is sketched by the gray dotted line.
Parameters are the same as in Fig.~\ref{fig:scaling} for $\gamma=10$.}
  \label{fig:tdse-duration}
 \end{figure}

\begin{figure}
   \includegraphics[width=8.5cm]{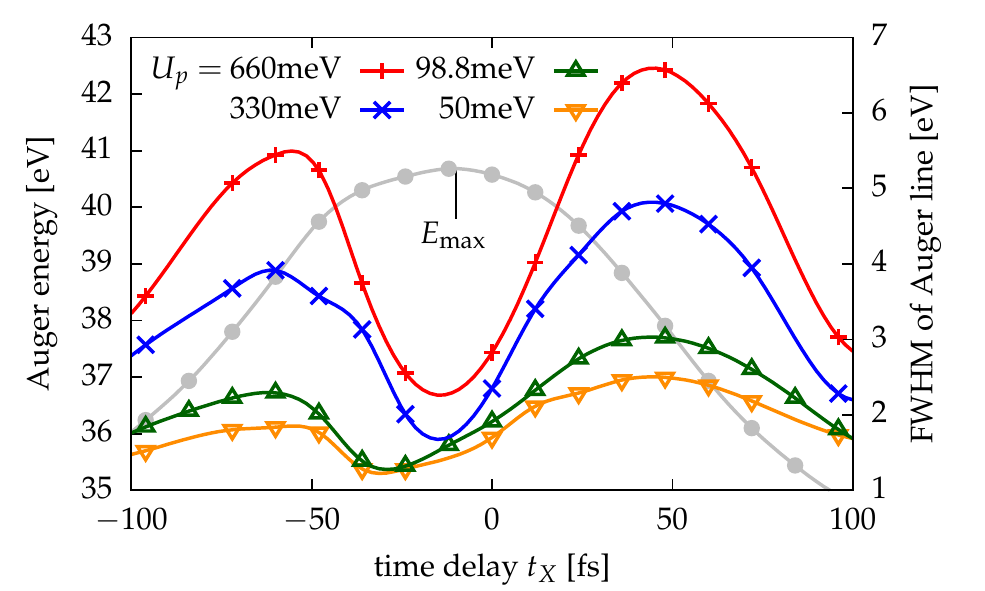}
   \caption{(color online) The same as Fig.~\ref{fig:tdse-duration} but for different ponderomotive potentials 
$U_p$ of the streaking field at a fixed XUV pulse duration of $28\,$fs ($\gamma=10$).
The graph of the maximum of the line corresponds to $U_p=98.8\,$meV.}
   \label{fig:tdse-up}	
\end{figure}

We can now analyze the dependence of the streaked lines upon various pulse parameters.
Those with most influence on the streaking mechanism are the ponderomotive potential $U_p$
 of the streaking field and the duration of the pump pulse, $\tau_X$.
In Fig.~\ref{fig:tdse-duration} the dependence of the Auger line width of the Xe \emph{NOO} transition
is shown for a set of XUV pulse durations for a scaling parameter of $\gamma=10$.
For larger pulse durations, a longer period of the slope of the streaking vector potential is
accessible, consequently leading to a larger overall width, which is in accordance with the
typical streaking mechanism. However, the asymmetry with respect to the sign of the slope of the vector 
potential is more pronounced for shorter pulse durations ($2.8\,$fs). This can be attributed to 
the fact, that for longer pulse durations, the line width is dominated by the streaking part and for 
shorter pulse durations the chirp becomes dominant, which will be discussed in detail in Sec.~\ref{sec:params}.

For different ponderomotive potentials of the streaking field, shown in Fig.~\ref{fig:tdse-up}, a similar
picture arises: the larger the ponderomotive potential, the larger is the streaking contribution leading to
a relative decrease of the observed asymmetry. However, we note that for very small $U_p$ the 
line width asymmetry must vanish because of the vanishing vector potential.

\subsection{Auger electron and photoelectron coincidence spectra}
\label{ssec:tdse-pe}
 \begin{figure}
  \includegraphics[width=8.5cm]{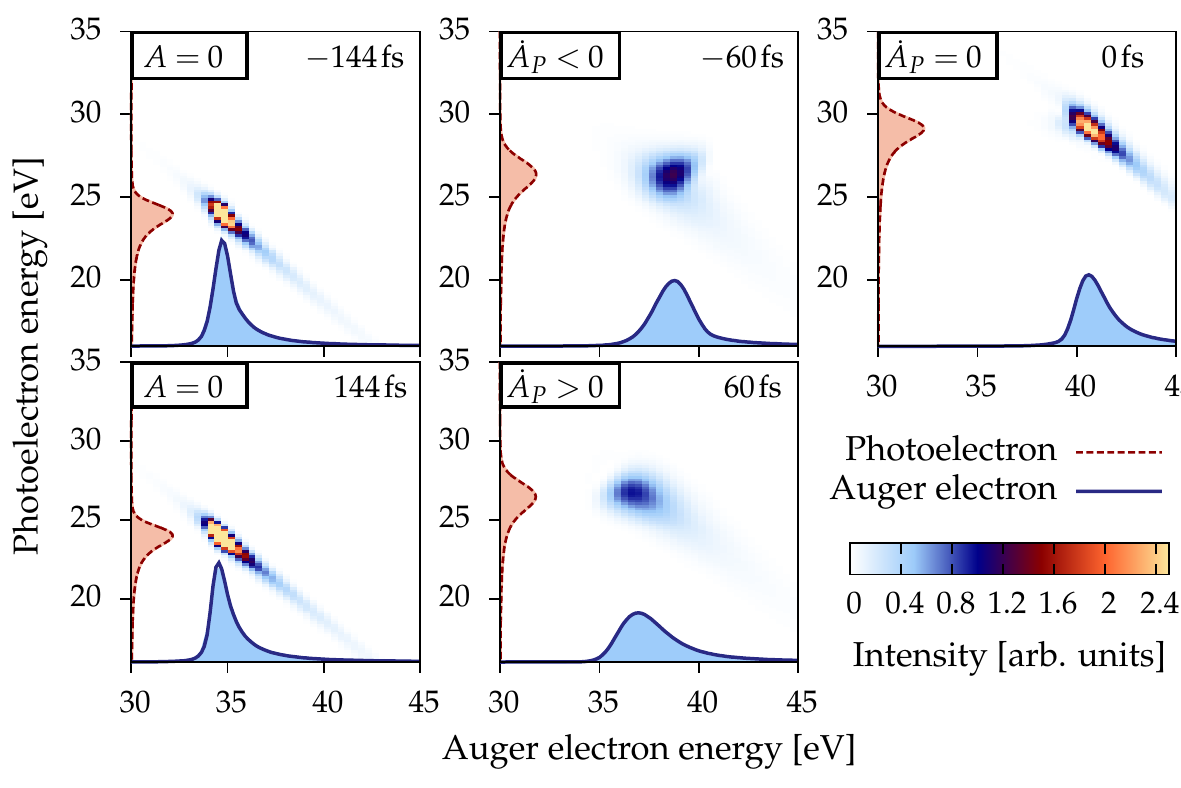}
  \caption{(color) Coincidence energy spectra for Auger and photoelectrons for selected time delays $t_X$ calculated by solving Eqs.~\eqref{eq:systdse-1} 
 and \eqref{eq:systdse-2}. The integrated photoelectron (Auger electron) distribution is plotted with red dashed (blue solid) lines. Parameters
are the same as in Fig.~\ref{fig:tdse-lines} (Xe \emph{NOO} with $\gamma=10$).}
  \label{fig:tdse-pe-ae}
 \end{figure}

Additionally to the individual kinetic energy spectra of the Auger electrons, Eqs.~\eqref{eq:systdse-1} and \eqref{eq:systdse-2} also allow
for the calculation of coincidence energy spectra of both involved electrons. This gives more detailed insight into their correlated motion.
 An example for the Xe \emph{NOO} transition ($\gamma=10$) is given in Fig.~\ref{fig:tdse-pe-ae} for five selected
time delays $t_X$.
All spectra are dominated by a diagonal line from top left to bottom right, which indicates an energy correlation between photoelectron and Auger electron. This is 
due to an energy exchange between both and governed by energy conservation. For the field-free cases ($\pm 144\,$fs) and the maximum of the
vector potential ($0\,$fs) a rather sharp spectrum is observed, whereas at the (approximate) zero transitions of the vector potential ($\pm 60\,$fs)
the streaking mechanism gives broad energy distributions, for both, the photoelectron and the Auger electron.

In addition to the integrated Auger electron spectra (blue, solid lines), the corresponding photoelectron 
distribution is plotted with (red) dashed lines. A careful inspection reveals, that the prominent asymmetry with respect to the
slope of the vector potential ($\pm 60\,$fs) observed for the Auger electron, is not present in the photoelectron spectra.
 This indicates, that the photoelectron distribution
carries no energetic chirp. A more detailed description and a simple picture for this are given in Sec.~\ref{ssec:pe-distribution}.

\section{Semi-classical simulations}
\label{sec:semi-classics}
To proceed further and compare quantitatively with current experiments, it
is crucial to take into account the 3D geometry of the atom and the true
time scales, i.e. to avoid the scaling procedure by $\gamma$.
 Since this is not possible utilizing TDSE simulations, it is necessary to
turn to a (semi-) classical description of FA-PCI including both electrons,
the ion and the streaking field.
Our classical method describing PCI is motivated by successful previous 
models for the field-free case \cite{niehaus_analysis_1977}.

The classical dynamics of both electrons is governed by Newton's equations ($m_e=1$),
\begin{eqnarray}
 \ddot{\boldsymbol{r}}_P(t) & = & \boldsymbol{F}_P(\boldsymbol{r}_P,\boldsymbol{r}_A,t)\;, \nonumber \\
 \ddot{\boldsymbol{r}}_A(t) & = & \boldsymbol{F}_A(\boldsymbol{r}_P,\boldsymbol{r}_A,t) \; ,
 \label{eq:newton}
\end{eqnarray}
where the photoelectron (Auger electron) is denoted by index P (A).
The propagation is split into two phases: (i) 
before Auger decay ($t_{P_i}\leq t<t_{A_i}$) and (ii) after Auger decay ($t_{A_i}<t<t_d$), where $t_d$ is the time at the detector, which determines
the corresponding forces in Eq.~\eqref{eq:newton}:
\begin{itemize}
 \item {\bf Phase (i)  [$t_{P_i} \leq t < t_{A_i}$]: } 
   \begin{equation}
     \boldsymbol{F}_P(t)=-\nabla V^+(\boldsymbol{r}_P) - \boldsymbol{E}_L(t)
     \label{eq:force_1}
   \end{equation}
 \item {\bf Phase (ii) [$t\geq t_{A_i}$]:}
   \begin{eqnarray}
     \boldsymbol{F}_P(t) & = & -\nabla V^{2+}(\boldsymbol{r}_P) - \boldsymbol{E}_L(t) - \nabla V_{e-e}(\boldsymbol{r}_A, \boldsymbol{r}_P), \nonumber \\
     \boldsymbol{F}_A(t) & = & -\nabla V^{2+}(\boldsymbol{r}_A) - \boldsymbol{E}_L(t) - \nabla V_{e-e}(\boldsymbol{r}_P, \boldsymbol{r}_A) .
     \label{eq:force_2}
   \end{eqnarray}
\end{itemize}
For (i) only the photoelectron is propagated in the combined field
of a singly charged ion, $V^+(\boldsymbol{r})$,  and the streaking field, $\boldsymbol{E}_L(t)$.
In phase (ii)  both electrons experience the potential of a doubly charged ion, $V^{2+}(\boldsymbol{r})$, 
the streaking field, and their binary interaction $V_{e-e}(\boldsymbol{r_1},\boldsymbol{r}_2)$.
All interaction potentials are of pure Coulomb type:
\begin{eqnarray}
 V^+(\boldsymbol{r})   =   -\frac{1}{|\boldsymbol{r}|}, \;   V^{2+}(\boldsymbol{r}) & =&-\frac{2}{|\boldsymbol{r}|} \\
  \textup{and} \;\;\; V_{e-e} (\boldsymbol{r}_1,\boldsymbol{r}_2)  &=&  \frac{1}{|\boldsymbol{r}_1-\boldsymbol{r}_2|} \; .
  \label{eq:potentials}
\end{eqnarray}
By setting $V_{e-e}\equiv 0$ and additionally considering $V^+\equiv V^{2+}$, all e-e interactions and PCI effects
can be turned off (denoted by ''neglecting e-e interaction`` in the following).
The set of equations~\eqref{eq:newton} is completed by associated initial conditions
\begin{eqnarray}
 \boldsymbol{\dot{r}}_P(t_{P_i}) & = & \boldsymbol{p}_{P_i} \;\; \textup{and} \;\;\; \boldsymbol{r}_{P}(t_{P_i})=\boldsymbol{r}_{P_i} \; ,
  \label{eq:initial-conditions-P} \\
 \boldsymbol{\dot{r}}_A(t_{A_i}) & = & \boldsymbol{p}_{A_i} \;\; \textup{and} \;\;\; \boldsymbol{r}_{A}(t_{A_i})=\boldsymbol{r}_{A_i} \; .
 \label{eq:initial-conditions-A}
\end{eqnarray}

\subsection{Initial condition sampling}
\label{ssec:initial-condition-sampling}
\begin{figure}
 \includegraphics[width=8.5cm]{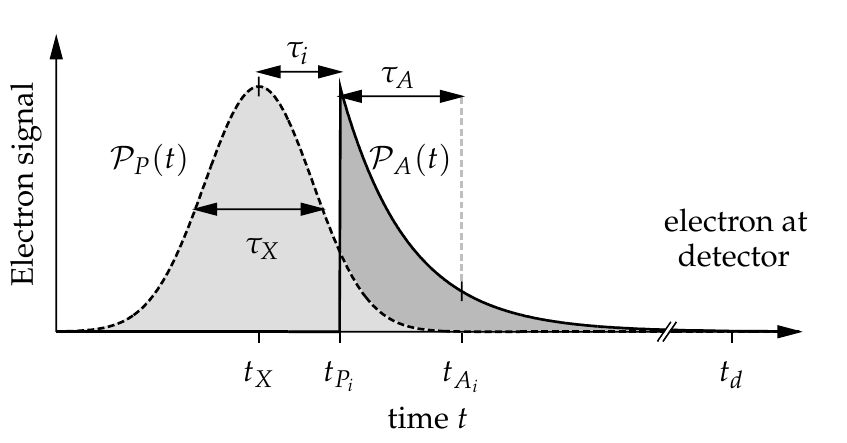}
 \caption{Temporal parameters and electron distributions: the XUV pulse is centered at $t_X$ with a FWHM
duration of $\tau_X$. At $t_{P_i}$ during the pulse the photoelectron
is excited which triggers Auger decay at a time instant $t_{A_i}$.
The measurement is performed long after the pulses and the decay are over, at time
 $t_d$ ($t_d \rightarrow \infty$).}
 \label{fig:times}
\end{figure}
To reproduce the quantum mechanical nature of photoionization and Auger decay in our classical model, we developed
a Monte Carlo (MC) sampling procedure for the initial conditions \eqref{eq:initial-conditions-P} 
and \eqref{eq:initial-conditions-A}.
During the XUV pulse, the photoelectron is released with the probability (proportional to the instantaneous intensity of the XUV pulse, $\propto E_X^2$)
\begin{equation}
 \mathcal{P}_{T_P}(\tau_i)=\frac{1}{\sqrt{\pi}\tau_X}\exp\left(-\frac{\tau_i^2}{\tau_X^2} \right )\; ,
 \label{eq:xuv-time-distribution}
\end{equation}
which creates the core hole at a time  $\tau_i=t_{P_i}-t_X$.
The vacancy is filled after the time $\tau_A=t_{A_i}-t_{P_i} > 0$ by lifting the Auger electron into the continuum
according to the decay law (probability density, see  Fig.~\ref{fig:times} for notations)
\begin{equation}
 \mathcal{P}_{T_A}(\tau_A)=\Gamma_A e^{-\Gamma_A \tau_A} \; .
 \label{eq:Auger-decay-law}
\end{equation}
The kinetic energy distribution of the photoelectron follows a Gaussian distribution,
\begin{equation}
 \mathcal{P}_{E_P}(E_{P_i})=\frac{1}{\sqrt{2 \pi}\sigma_X} \exp \left (- \frac{({E_{P_i}}-E_{P_0})^2}{2 \sigma_X^2}\right )\;,
 \label{eq:photoelectron-energy}
\end{equation}
with the spectral width $\sigma_X$ centered around the energy 
$E_{P_0}=\omega_X-I_p$, with the ionization potential of the core electron  $I_p$.
The (undisturbed) line shape of the Auger electron with mean energy $E_A$ associated with Eq.~\eqref{eq:Auger-decay-law} is a Lorentzian distribution
\begin{equation}
 \mathcal{P}_{E_A}(E_{A_i})=\frac{\Gamma_A/2\pi}{ (E_{A_i}-E_{A})^2+\frac{1}{4}\Gamma_A^2} \; .
 \label{eq:Auger-energy}
\end{equation}
With that, the absolute values of the initial momenta are set by Eqs.~\eqref{eq:photoelectron-energy} and
\eqref{eq:Auger-energy} to
\begin{equation}
 |\boldsymbol{p}_{P_i}|=\sqrt{2 E_{P_i}} \;\; \textup{and} \;\;  |\boldsymbol{p}_{A_i}|=\sqrt{2 E_{A_i}} \; .
 \label{eq:initial-momenta}
\end{equation}
For small initial distances $r_{P_i}$ and $r_{A_i}$ of the electrons from the ion, it is important to take into account the remaining
finite binding potential at the point of appearance of the electrons, 
$V^+(\boldsymbol{r}_{P_i})$ and $V^{2+}(\boldsymbol{r}_{A_i})$,
 to assure their correct asymptotic momenta on the detector.
Entering as a free parameter in our model, we carefully checked the influence of different values of $r_{P_i}$ and $r_{A_i}$
ranging from $1$ to $20$ (in units of the Bohr radius) and found no significant change of the results.

The directions of  $\boldsymbol{r}_{P_i}$ and $\boldsymbol{r}_{A_i}$ as well as of $\boldsymbol{p}_{P_i}$ and $\boldsymbol{p}_{A_i}$
 are given by the quantum mechanical angular distributions of the associated initial state, approximated by
\begin{equation}
 \mathcal{P}_{P/A}(\varphi)=\frac{1}{4\pi}\left[1+ \beta_{P/A} (3\cos^2 \varphi -1) \right]
 \label{eq:angular-distribution}
\end{equation}
with the asymmetry parameter $\beta$, being available in the literature, e.g. \cite{snell_angular_2000}.
We note as a technical aspect, that sphere point picking \cite{wolfram_sphere} is crucial for the correct MC sampling of 
Eq.~\eqref{eq:angular-distribution} to maintain the correct uniform distribution of points on a sphere.

\subsection{Extraction of observables}
We propagate Eq.~\eqref{eq:newton} with initial conditions \eqref{eq:initial-conditions-P} and 
\eqref{eq:initial-conditions-A}, randomly distributed according to Eqs.~(\ref{eq:xuv-time-distribution}-
\ref{eq:angular-distribution}),  utilizing a velocity Verlet algorithm
with an adaptive time step size control, see e.g. \cite{ott_md_2010}.
This method will be called ''Monte-Carlo Molecular Dynamics`` (MC-MD) simulations in the following
(MD refers to the classical propagation of both interacting electrons leaving the atom).

For each run, the final momenta $\boldsymbol{p}_{P_f}$ and $\boldsymbol{p}_{A_f}$
of typically $10^6$--$10^7$ trajectories are recorded and sorted in angle- and energy-resolved
histograms until convergence is reached.
The Auger electron kinetic energy spectra are then obtained by integrating over a detector angle element 
of $12.5^\circ$, typical for experiments, around the field polarization axis $\hat{e}_z$. 
Two opposite detection directions 
are possible, determined by the direction of $\boldsymbol{A}$. We will only show results for 
the detector with positive energy shift at the maximum of the single-cycle vector potential; 
the second detector gives the same results, but for changed sign
in $\boldsymbol{A}$. In experiments it is often favorable to consider two opposing detectors
to assure the same streaking conditions \cite{schuette_evidence_2011}.
Post-processing of the Auger line shapes is performed similar to the TDSE case, cf. Sec.~\ref{ssec:tdse-results}.
Additionally, as in the previous part, we restrict ourselves to the case of Auger electrons,
the analysis of the photoelectrons can be performed in a similar way.

\subsection{MC-MD-Results}
\begin{figure}
 \includegraphics[width=8.5cm]{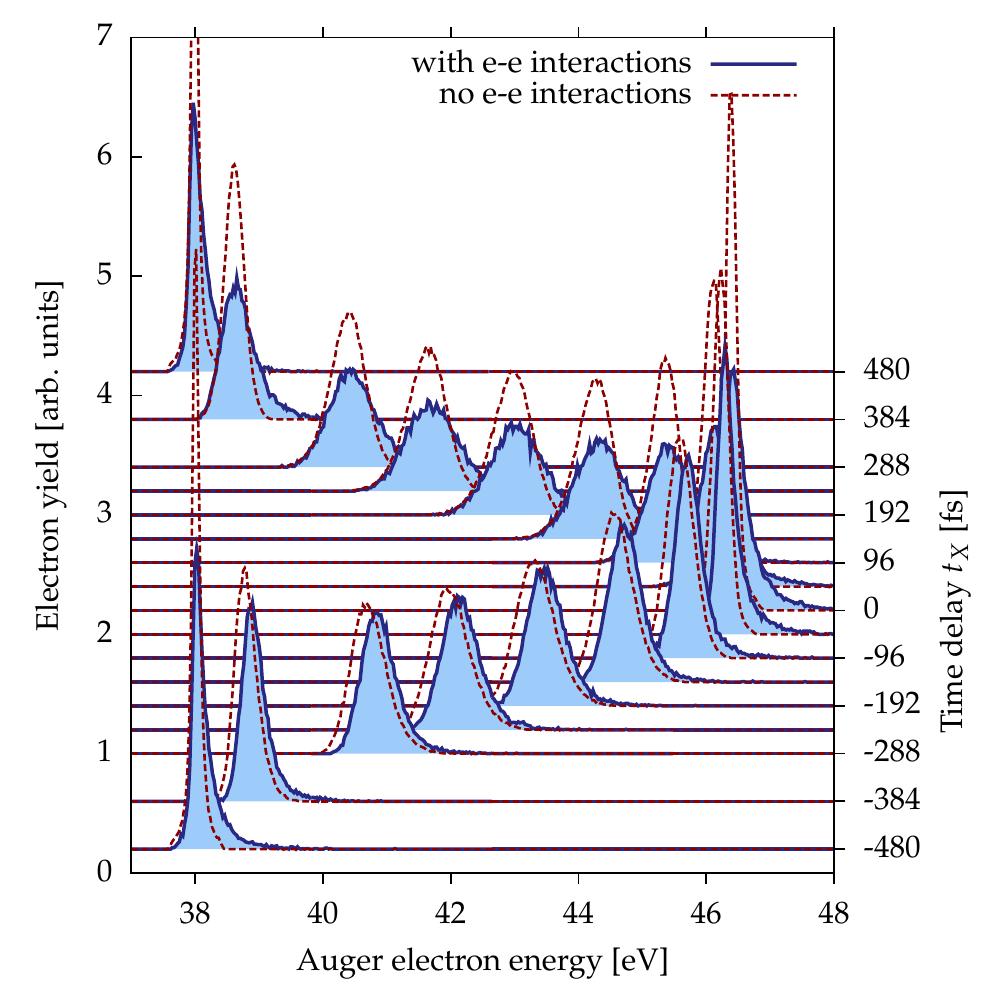}
 \caption{(color online) Auger line shapes of the Kr \emph{MNN} transition  for a set of time delays $t_X$ in a $1\,$THz 
single cycle streaking field with $U_p=80\,$meV for a XUV pulse duration of  $28\,$fs at a photon energy of $97\,$eV.
Results are obtained by MC averaging of MD trajectories. Shown is the case including photoelectron-Auger electron interactions (blue
solid lines) and neglecting e-e interactions (red dashed lines). Note: the latter lines are shifted by $100\,$meV towards higher energy for clarity.  }
 \label{fig:md-lines}
\end{figure}
We may now drop the scaling procedure \eqref{eq:scaling} introduced for TDSE simulations and restore the 
true time constants. The result for a 
full scan of time delays for the Krypton \emph{MNN} transition in a $1\,$THz streaking field with
a ponderomotive potential of $80\,$meV is shown in Fig.~\ref{fig:md-lines} for the cases (i) including 
(blue solid lines) and (ii) neglecting (red dashed lines) e-e interactions. A similar picture as for the TDSE
simulations, cf. Fig.~\ref{fig:tdse-lines}, arises. A prominent asymmetry with respect to positive
and negative time delays, i.~e.~$\dot{A}(t)|_{t=t_X} \equiv \dot{A}_P>0$ and $\dot{A}_P<0$ respectively, can be found for (i) which 
completely vanishes for (ii). Note: again, the lines with PCI effects excluded are shifted towards
higher energy by $100\,$meV for better
comparison. The shift is smaller compared to Fig.~\ref{fig:tdse-lines} due to the fact that $\gamma=10$
overestimates PCI in the case of the TDSE simulations.

\begin{figure}
 \includegraphics[width=8.5cm]{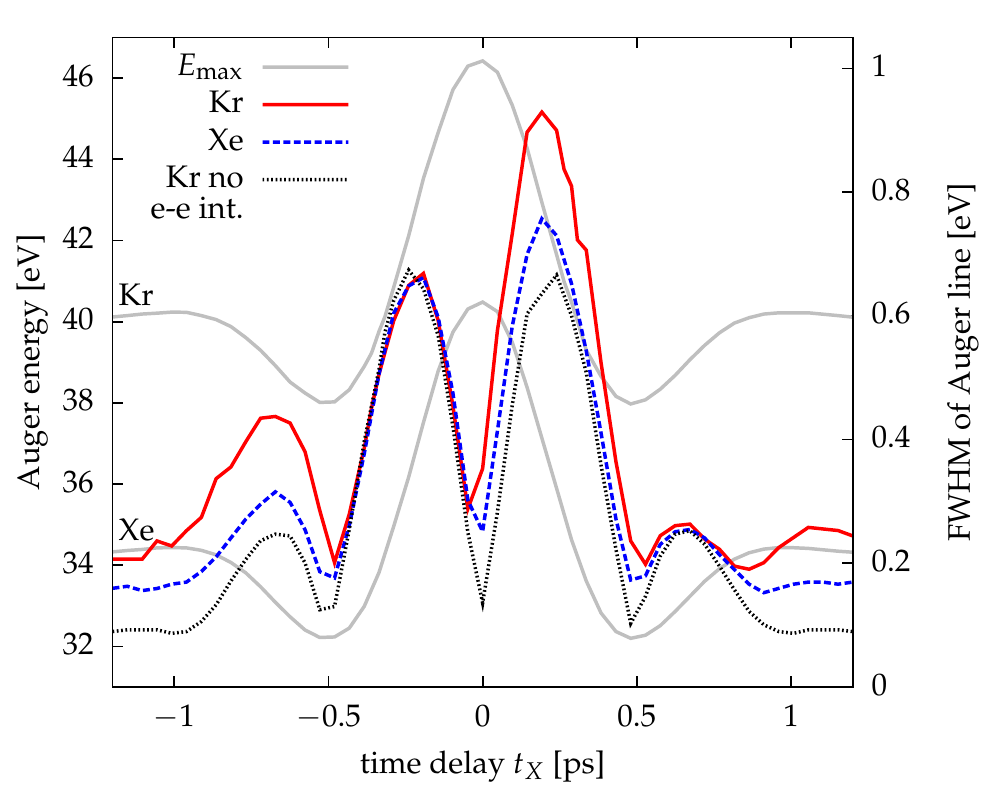}
 \caption{(color online) FWHM of the Auger lines (right axis) of the Xe \emph{NOO} and Kr \emph{MNN} transitions calculated 
utilizing MC-MD simulations.
For Xe (Kr) the photon energy of $91\,$eV ($97\,$eV) leads to a photoelectron kinetic energy of $24\,$eV ($2.6\,$eV).
All other parameters are the sames as in Fig.~\ref{fig:md-lines}. The position of the maximum of the line, $E_{\textup{max}}$ for
 both cases is shown in gray. The case neglecting e-e interactions for Kr is given by the black dotted line.}
\label{fig:md-Kr-Xe}
\end{figure}

The FWHM and position of the line for the Xe \emph{NOO} and Kr \emph{MNN} decays are shown in Fig.~\ref{fig:md-Kr-Xe}.
At the considered photon energies of $91\,$eV for the former and $97\,$eV for the latter, photoelectron energies
of $24\,$eV and $2.6\,$eV at comparable Auger electron energies of $34\,$eV and $40\,$eV are observed.
Due to the slow photoelectron, for Kr a dramatic increase of PCI in comparison to Xe is expected, 
which is connected with a stronger chirp on the Auger electron's energy.
 This is confirmed by our calculations (red solid lines  vs. blue dashed lines). 
If e-e interactions are neglected, similar line shapes and widths are observed for rising and falling
flank of the vector potential (black dotted line). These observations are in qualitative agreement with 
TDSE simulations discussed in Fig.~\ref{fig:tdse-pevel} and confirm that PCI is the origin for the Auger electron's chirp.

\subsection{Comparison with TDSE}
\begin{figure}
 \includegraphics[width=8.5cm]{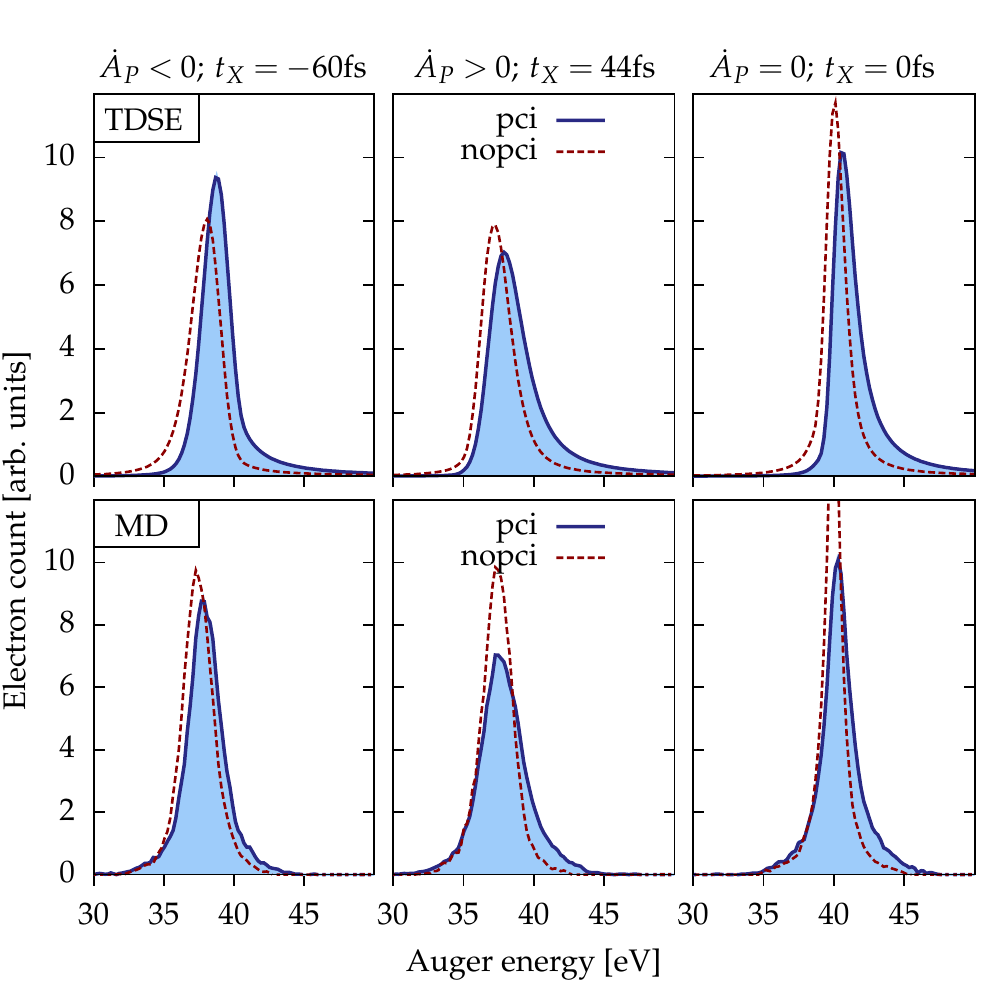}
 \caption{(color online) Line shapes of the Xe \emph{NOO} transition ($\gamma=10$) at  three different
time delays $t_X$, corresponding to the FWHM maxima (see curve in Fig.~\ref{fig:tdse-pevel} for $91\,$eV) and $t_X=0$, for
 streaking with $U_p=98.8\,$meV. Results from TDSE simulations (top row) and MC averaged MD simulations
(bottom row) are given. The cases including PCI (blue solid lines)
and neglecting PCI (red dashed lines) are compared for falling (left), rising (center) and zero
 (right) slope of $A(t)$.}
 \label{fig:tdse-3dmd-comparison}
\end{figure}

By construction, the MD simulations neglect any quantum effects in the electron dynamics, such as coherence, interference and spin.
To test the above-introduced technique, a detailed comparison of the line shapes
calculated utilizing MD and TDSE methods for three different
time delays is presented in  Fig.~\ref{fig:tdse-3dmd-comparison}.
The Auger electron spectrum of the Xe \emph{NOO} transition, necessarily scaled by a factor of $\gamma=10$ for 
both simulations, is given for $\dot{A}_P<0$ (left), $\dot{A}_P>0$ (center) and $\dot{A}_P=0$ (right)
for situations including (solid lines) and neglecting (dashed lines) PCI.
For better comparison, the Auger spectra obtained from TDSE and MD simulations have both been renormalized.
This rescaling is necessary due to the 
small XUV ionization cross section, which has been neglected in the classical simulations.

As a first observation, the line shapes obtained by MD simulations (bottom) are slightly broadened in comparison 
to the TDSE (top). This can be attributed to the averaging over the finite detector
acceptance angle of $12.5^\circ$ in the 3D MD calculations. Here, trajectories are collected, which 
have been streaked with smaller amplitude due to their initial deviation (angular distribution) from 
the field-polarization axis.
For both types of simulations, the line for $\dot{A}_P>0$ is significantly broader than for $\dot{A}_P<0$
which completely vanishes if PCI is turned off. Furthermore, both methods reproduce a similar PCI-induced
shift of the line to higher energies. Thus the general trends as well as the underlying mechanism
for the description of the asymmetry are correctly captured by the MD model and quantum effects in the electron propagation play
no dominant role for the line width in the considered excitation regimes.

\section{Analytical model for Auger line shapes }
\label{sec:analytic}
In the previous sections, we have shown, utilizing TDSE and MD simulations,
that Auger emission is chirped if PCI is involved which has a prominent impact on the line shapes
in external laser fields.
To get deeper insight in the underlying physics, we derive closed expressions
for the line shape of the Auger electron in the streaking field including
PCI effects based on a classical 1D model. 

\subsection{Time-to-energy mapping}
The key mechanism of streaking is the mapping between a temporal process and the measurable 
energy or momentum distribution. 
For Auger electrons, the temporal distribution follows the decay law, Eq.~\eqref{eq:Auger-decay-law}.
The corresponding probability to find the Auger electron in the continuum at a time $t$ is given by
\begin{equation}
 \mathcal{P}_A(t) = \int_{0}^t \: \textup{d}\tau_A \mathcal{P}_{T_A}(\tau_A) \; ,
 \label{eq:time-distribution}
\end{equation}
which approaches unity for long times (see Sec.~\ref{ssec:initial-condition-sampling} and Fig.~\ref{fig:times}
for notations).
The distribution \eqref{eq:time-distribution} is translated by the streaking field to energy, thus
the quantity of interest is the kinetic energy change of the Auger electron measured at a remote detector
at time $t_d$. Its final momentum is given by $p_{A_i}+\Delta p(t_{A_i})$. The field-induced
momentum change evaluates to 
\begin{equation}
 \Delta p=-\int_{t_{A_i}}^{t_d} \textup{d}\overline{t} \; E(\overline{t}) = -A(t_{A_i}) \;,
\end{equation}
where vanishing of the vector potential for $t = t_d$ with $t_d \rightarrow \infty$ is assumed.
With that, we obtain for the Auger electron energy change
\begin{equation}
 E_{\textup{kin}}^d=\frac{1}{2}\left[p_{A_i}-A(t_{A_i})\right]^2+\Delta E^{\textup{PCI}}-\frac{p_{A_i}^2}{2}\; .
 \label{eq:kinetic_energy_change}
\end{equation}
A possible energy exchange between photoelectron and Auger electron due to post-collision interaction
is accounted for by $\Delta E^{\textup{PCI}}(t_X,\tau_A)$.
It depends on the distance from the ion, i.~e. on the Auger time delay $\tau_A$, and the pump-probe time delay $t_X$.
In the following we consider fixed (sharp) initial momenta of the two electrons, $p_{P_i}$ and $p_{A_i}$.

Let us first assume an infinitesimal duration of the pump pulse ($\tau_X \rightarrow 0$), which
corresponds to $t_{P_i}\equiv t_X$. An extension of the model to finite XUV pulse durations will be 
presented in Sec.~\ref{ssec:finite-xuv}.
Expanding $A$ around $t_{P_i}$  to second order gives for the $\tau_A$-dependent energy shift
 $\epsilon_S^\textup{PCI}\equiv E_{\textup{kin}}^d+p_{A_i}A_P$:
\begin{equation}
 \epsilon_S^{\textup{PCI}} \approx -p_{A_i} \left ( \dot{A}_P \tau_A + \frac{1}{2} \ddot{A}_P\tau_A^2 \right )+\Delta E^{\textup{PCI}} (\tau_A)\;.
 \label{eq:epsilon}
\end{equation}
Here, we use the notations $A_P\equiv A(t_{P_i})$, $\dot{A}_P=\partial/\partial t A(t) |_{t=t_{P_i}}$ and 
$\ddot{A}_P=\partial^2/\partial t^2 A(t)|_{t=t_{P_i}}$ and neglect higher-order
 terms $\mathcal{O}\left(\tau_A^3 \omega_L^3,A^2\right)$.
Eq.~\eqref{eq:epsilon} translates the temporal distribution of Auger electrons governed by Eq.~\eqref{eq:time-distribution} 
to the energy domain through action of the streaking vector potential and PCI. 
 This procedure was first applied in \cite{ogurtsov_auger_1983} for the time-to-energy transformation
due to PCI without external fields.
In the present paper, we demonstrate, extending the simplified model of Ref.~\cite{schuette_evidence_2011},
 this mapping including PCI and streaking, which
 gives direct access to closed expressions for the Auger line shape of FA-PCI.
\subsection{Lineshapes neglecting PCI}
Let us first consider the case $\Delta E^{\textup{PCI}}=0$ in Eq.~\eqref{eq:epsilon} and find
the Auger line shapes at characteristic pump-probe time delays $t_X$ for zero transitions and maxima of $A(t)$.
\subsubsection{Zero transitions of the vector potential.}
\begin{figure}
 \includegraphics[width=4.25cm]{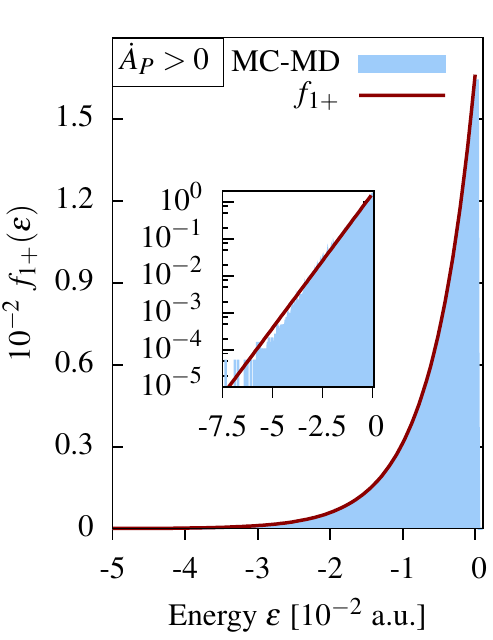}
 \includegraphics[width=4.25cm]{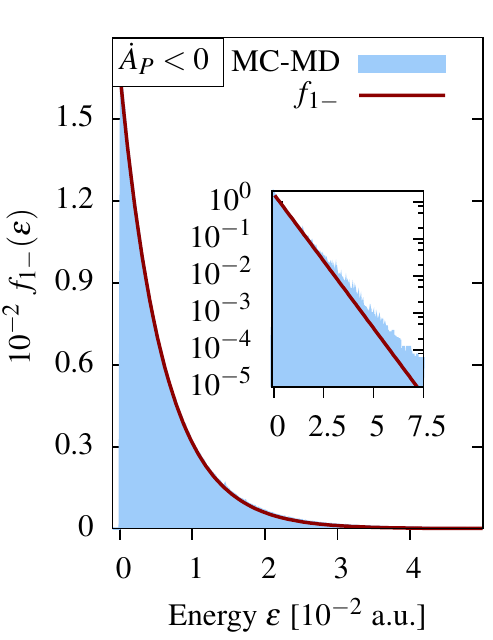}
 \caption{(color online) Analytical model for the Auger line shape neglecting PCI, Eq.~\eqref{eq:lineshape-nopci-azero}, 
compared to corresponding MD simulations (excluding PCI and sampling of $p_{A_i}$)
 at zero transitions of the 
vector potential for increasing (left) and decreasing (right) slope of $A$.
Shown is Xe $\emph{NOO}$ in  a $1\,$THz streaking field with a duration of $1\,$ps and a ponderomotive potential
of  $100\,$meV. The insets show the same data semi-logarithmically.}
 \label{fig:lineshape-nopci-azero}
\end{figure}

 Since $\dot{A}_P\neq 0$, the leading contribution to the mapping~\eqref{eq:epsilon} is linear in $\tau_A$ and higher-order
contributions can be dropped, which gives
\begin{equation}
 \tau_A=-\frac{\epsilon}{p_{A_i} \dot{A}_P} \;.
 \label{eq:tauA-nopci-Azero}
\end{equation}
Substituting expression \eqref{eq:tauA-nopci-Azero} in Eq.~\eqref{eq:time-distribution} 
gives the Auger line shapes for increasing (+) and decreasing (-) slope of $A$,
\begin{equation}
 f_{1\pm}(\epsilon)=\Gamma_1 e^{\pm \Gamma_1 \epsilon}, \;\;\; \textup{with} \;\;\; \Gamma_1=\frac{\Gamma_A}{p_{A_i}|\dot{A}_i|},
 \label{eq:lineshape-nopci-azero}
\end{equation}
with the normalization conditions
\begin{equation}
 \int_{-\infty}^{0}\textup{d}\epsilon  \;f_{1+}(\epsilon)=1,\;\;\; \textup{and} \;\;\; \int_{0}^{\infty} \textup{d} \epsilon\; f_{1-}{(\epsilon)} =1 \; .
\end{equation}
The comparison with 1D MD simulations for $\tau_X \rightarrow 0$, neglecting PCI and without sampling of the initial momentum $p_{A_i}$,
is shown in Fig.~\ref{fig:lineshape-nopci-azero} for Xe \emph{NOO} decay in a $1\,$THz streaking field.
The streaked lines exhibit
the same exponential decay law as the time dependence of the core hole decay.
 The direction of the slope of $A$ only affects the orientation
of the exponential tail. Deviations of Eq.~\eqref{eq:lineshape-nopci-azero} from the numerical 
solution are very small and are due to the linearization
of $A$ and are only visible in the logarithmic 
representation (insets in Fig.~\ref{fig:lineshape-nopci-azero}).

\subsubsection{Extrema of the vector potential}
\begin{figure}
 \includegraphics[width=4.25cm]{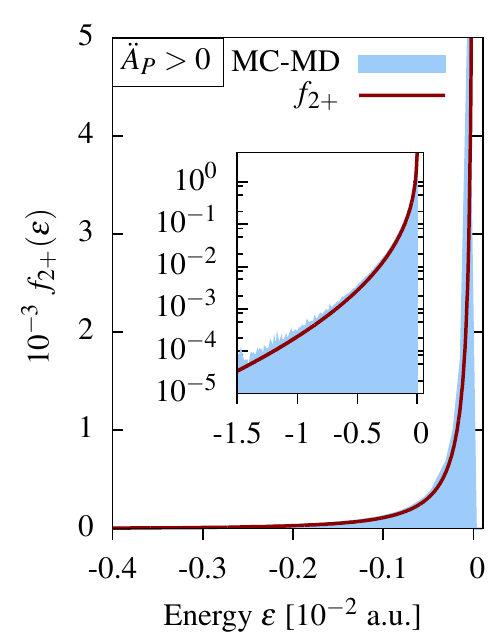}
 \includegraphics[width=4.25cm]{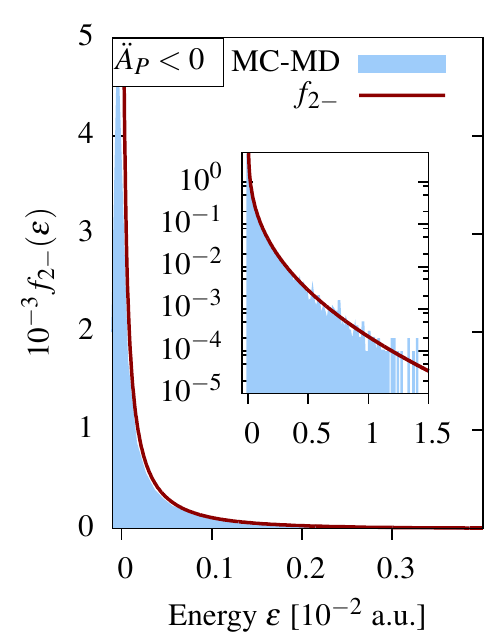}
 \caption{(color online) Auger lineshape for minima (left panel, $\ddot{A}_P>0$) and maxima 
(right panel, $\ddot{A}_P<0$) of $A$. 
Displayed is Eq.~\ref{eq:lineshape-nopci-aext} compared to MD simulations.
Parameters are the same as for Fig.~\ref{fig:lineshape-nopci-azero}.}
 \label{fig:lineshape-nopci-aext}
\end{figure}

For maxima ($\ddot{A}_P<0$) and minima ($\ddot{A}_P>0$) of the vector potential, 
$\dot{A}_P=0$ holds, thus the second order in $\tau_A$ is the leading contribution in
Eq.~\eqref{eq:epsilon}.
Because obviously $\tau_A \geq 0$, we obtain only one solution
\begin{equation}
 \tau_A=\sqrt{-\frac{2\epsilon}{\ddot{A}_P p_A}}\; .
 \label{eq:tauA-nopci-Amax}
\end{equation}
The corresponding line shapes for maxima (``-'') and minima (``+'') of $A$ evaluate to
\begin{eqnarray}
 f_{2\pm}(\epsilon)=\frac{\Gamma_2}{2} \frac{1}{\sqrt{|\epsilon|}} e^{-\Gamma_2 \sqrt{|\epsilon|}},\;\; \Gamma_2=\Gamma_A \sqrt{\frac{2}{p_{A_i |\ddot{A}_P|}}}\; ,
 \label{eq:lineshape-nopci-aext}
\end{eqnarray}
with the normalizations
\begin{eqnarray}
 \int_{-\infty}^{0} \textup{d}{\epsilon}\; f_{2+}(\epsilon)=1,\;\;\; \textup{and} \; \int_{0}^{\infty} \textup{d} \epsilon\;f_{2-}(\epsilon) = 1 \; .
\end{eqnarray}
The comparison with MD data is given in Fig.~\ref{fig:lineshape-nopci-aext}. The line is
dominated by a sharp onset at zero and a rather rapid decay. 
The second order expansion of $A$ gives perfect agreement with
the simulation (logarithmic representation given in inset in Fig.~\ref{fig:lineshape-nopci-aext}).

\subsection{Analytical model for PCI}
\label{ssec:lineshapes-pci}
\begin{figure}
 \includegraphics[width=8.5cm]{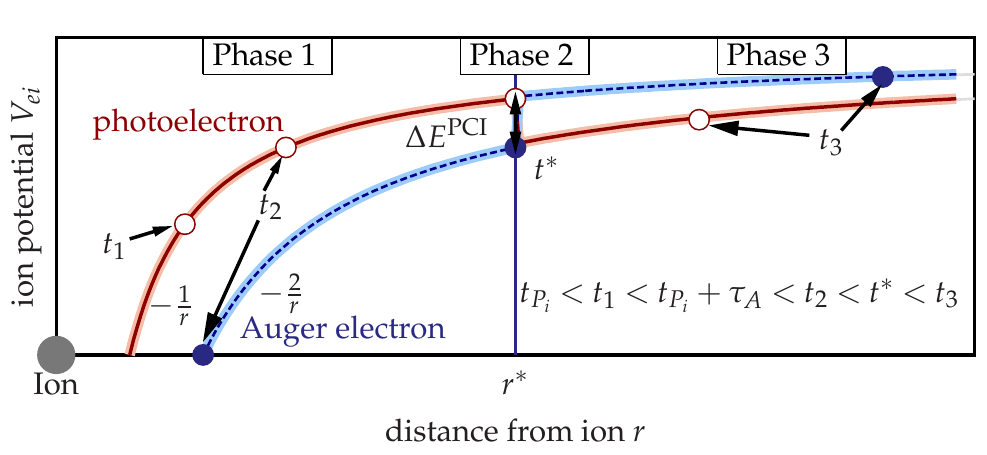}
 \caption{(color online) Scheme of simplified 1D propagation with PCI effects: in phase 1 ($t<t^*$), 
the Auger electron (filled circles) catches up with the slow photoelectron (open circles) and overtakes it at
$t=t^*$ (phase 2), $p_A>p_P$ is assumed. The propagation towards the detector  in phase 3 ($t>t^*$) is similar to
phase 1. Figure after ref.~\cite{schuette_evidence_2011}.}
 \label{fig:pcimodel}
\end{figure}
We now consider the case $\Delta E^{\textup{PCI}}\neq 0$ in Eq.~\eqref{eq:epsilon}.
To obtain closed expressions for the lineshapes including streaking and PCI, the (semi-)classical model introduced
in Sec.~\ref{sec:semi-classics} needs to be simplified in order to calculate $\Delta E^{\textup{PCI}}$.
Following \cite{ogurtsov_auger_1983,russek_post-collision_1986},
 we neglect the direct electron-electron interaction and model the PCI energy exchange by an 
instantaneous change in the ionic binding potential of $X^{2+}$ to $X^{+}$ for the Auger electron and $X^{+}$ to $X^{2+}$ 
for the photoelectron. 
The propagation scheme, sketched in Fig.~\ref{fig:pcimodel}, then reads as follows:
\begin{itemize}
 \item {\bf Phase 1 ($t<t^*$):}  propagation of the photoelectron ($t>t_{P_i}$) and the Auger electron ($t>t_{A_i}$) without interaction in the streaking field $E_L(t)$:
 \begin{equation}
    \ddot{r}_P=-E_L(t) \; \textup{and} \; \ddot{r}_A=-E_L(t)
 \label{eq:eom_phase1}
 \end{equation}
 with initial conditions
\begin{eqnarray}
  r_{P}(t_{P_i})=r_{P_i},\;&&\; p_P(t_{P_i})=p_{P_i} \nonumber \\
  r_A(t_{A_i})=r_{A_i}, \;&&\;   p_A(t_{A_i})= p_{A_i}
 \label{eq:eom_phase1_ic}
\end{eqnarray}
 \item{\bf Phase 2 ($t=t^*$):} The Auger electron overtakes the photoelectron, changed screening of the ion's charge leads to
 energy exchange $\pm \Delta E^{\textup{PCI}}=1/r^*$
 corresponding to a momentum change of
 \begin{eqnarray}
  p_P \rightarrow p_P^-&=&p_P+\Delta p_P(\Delta E^{\textup{PCI}},t^*) \nonumber \\
  p_A \rightarrow p_A^+&=&p_A+\Delta p_A(\Delta E^{\textup{PCI}},t^*)
 \label{eq:eom_phase2}
 \end{eqnarray}
 \item{\bf Phase 3 ($t>t^*$):} similar to phase 1 but with initial conditions
 \begin{eqnarray}
   r_P(t^*)=r_A(t^*)=r^*; \nonumber \\
   p_P(t^*)=p_P^-\; \textup{and} \; p_A(t^*)=p_A^+ \; .
   \label{eq:eom_phase3_ic}
 \end{eqnarray}
\end{itemize}
A straightforward integration of Eq.~\eqref{eq:eom_phase1} gives the time of overtaking 
\begin{equation}
 t^*  =  t_{A_i} + \frac{\tau_A \tilde {p}_P - r_{AP}}{p_{AP}} \; ,
\label{eq:tstar}
\end{equation}
and the corresponding distance from the ion
\begin{equation}
  r^*= \tau_{A}p_r + \delta r^* \; .
  \label{eq:rstar}
\end{equation}
Here, we introduced the notations
\begin{eqnarray}
 \tilde{p}_P \equiv p_{P_i}-A(t_{P_i}) \;, \nonumber\\
 \tilde{p}_A \equiv p_{A_i}-A(t_{A_i}) \;, \nonumber
\end{eqnarray}
and
\begin{eqnarray}
 r_{AP}& \equiv&r_{A_i} - r_{P_i} - \int_{t_{P_i}}^{t_{A_i}} \; \textup{d}\tilde{t} A(\tilde{t}) \; , \nonumber \\[0.3pc]
 p_{AP}&\equiv&\tilde{p}_A-\tilde{p}_P=p_{A_i}-p_{P_i}+A(t_{P_i})-A(t_{A_i}) \;, \nonumber \\[0.3pc]
 p_r &\equiv& \frac{p_{A_i}p_{P_i}}{p_{AP}} \;, \nonumber\\ [0.3pc]
 \delta r^* &\approx& \frac{\tilde{p}_A r_{P_i}-\tilde{p}_P r_{A_i} - A(t_{A_i})(r_{A_i}-r_{P_i})}{p_{AP}} \; . \nonumber
\end{eqnarray}
With that, we obtain from Eq.~\eqref{eq:epsilon} the $\tau_A$ dependence of the 
time-to-energy mapping function including PCI and streaking:
\begin{equation}
 \epsilon^{\textup{PCI}}(\tau_A)=-p_{A_i} \dot{A}_p \tau_A + \frac{1}{p_r \tau_A +\delta r^*} \; .
  \label{eq:epsilon-pci}
\end{equation}
The distance $\delta r^*$ depends on the initial coordinates of the two electrons and their field-changed initial momenta. In most cases 
$\delta r^*$ will be a small correction to $r^*$. However, for situations with slow photoeletrons and relatively fast
Auger electrons, i.~e. situations with strong PCI, $\delta r^*$ may become large.
In the following, we derive generalized Auger line shapes for FA-PCI, improving the results presented in \cite{schuette_evidence_2011}, where $\delta r^*=0$
was assumed.
\subsection{FA-PCI without streaking}
\begin{figure}
 \includegraphics[width=6cm]{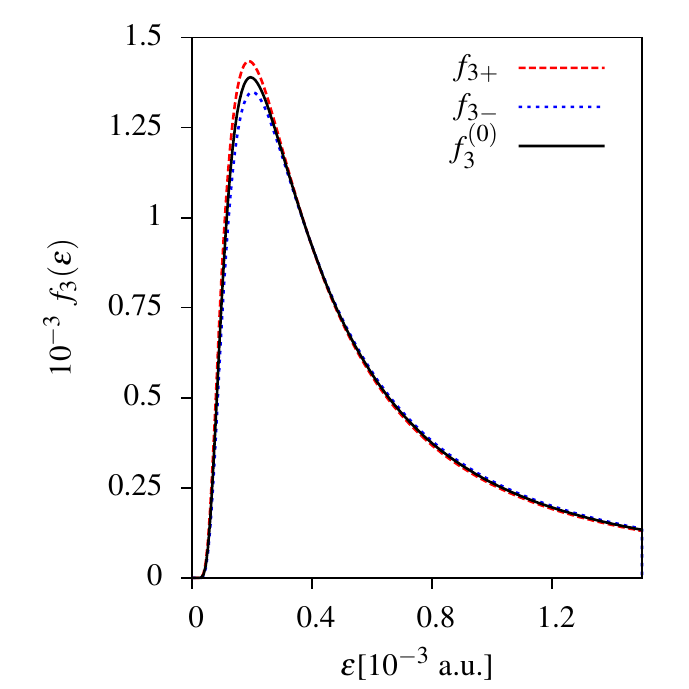}
 \caption{(color online) Field-assisted PCI without streaking contribution at zero transitions of $A$.
Shown is Eq.~\eqref{eq:lineshape-pci-nostreak} for $\dot{A}>0$ (red dashed line) and $\dot{A}<0$ (blue dotted line)
in comparison to the field free case, Eq.~\eqref{eq:lineshape-pci-nofield}. 
Data is for \emph{N-OO} transition of Xe in a $1\,$THz streaking field with $U_p=100\,$meV for $\delta r^*=1$.}
 \label{fig:pci-only}
\end{figure}
Before using the full mapping, let us neglect the first term in Eq.~\eqref{eq:epsilon-pci}
 linear in $\tau_A$ that is attributed to the 
streaking contribution discussed before.
Then we have a hyperbolic mapping function
\begin{equation}
 \epsilon_0^{\textup{PCI}} = \frac{1}{p_r \tau_A + \delta r^*} \;.
 \label{eq:epsilon-pci-only}
\end{equation}
Utilizing Eq.~\eqref{eq:epsilon-pci-only}, the straightforward transformation 
of the time distribution \eqref{eq:time-distribution} gives for the 
PCI-induced energy change of the Auger line shape
\begin{eqnarray}
 f_{3\pm}(\epsilon) = \Gamma_3 \frac{1}{\epsilon^2} e^{\Gamma_3 (\delta r^*-1/\epsilon )},\;\;\; \textup{with} \; \Gamma_3=\frac{\Gamma_A}{p_r}
\label{eq:lineshape-pci-nostreak}\\
 \int_{0}^{1/\delta r^*} \textup{d}\epsilon\; f_3(\epsilon) =1 \;,
\end{eqnarray}
where ``+'' (``-'') refers to $\dot{A}_P>0$ ($\dot{A}_P<0$) in $p_r$ and $\delta r^*$.
From Eq.~\eqref{eq:lineshape-pci-nostreak} we can immediately read off the energy distribution for the field-free case (and assuming $\delta r^*=0$),
\begin{equation}
 f_3^{(0)} (\epsilon) = \frac{\Gamma_A}{p_r^0} \frac{1}{\epsilon^2} e^{-\Gamma_A/(p_r^0 \epsilon)}\; ,
 \label{eq:lineshape-pci-nofield}
\end{equation}
with $p_r^0=p_{P_i} p_{A_i}/(p_{P_i}-p_{A_i})$, in accordance with the result given in \cite{ogurtsov_auger_1983}.
Our result \eqref{eq:lineshape-pci-nostreak} differs in the way, that although we exclude the explicit streaking contribution in Eq.~\eqref{eq:epsilon-pci},
the field-changed initial momenta $\tilde{p}_P$ and $\tilde{p}_A$ are included.
An example of Xe \emph{NOO} in a $1\,$THz streaking field is shown in Fig.~\ref{fig:pci-only}. 
In addition to the case of positive and negative slope of $A$, the field-free case, Eq.~\eqref{eq:lineshape-pci-nofield},
 is displayed.
For this specific case, no strong influence of the field on the pure PCI process is visible.
However, $f_{3+}$ has a slightly higher maximum, corresponding 
to smaller width, in contrast to the effect observed in the simulations in the previous section (note: this 
asymmetry is not the observed chirp). 
$f^{(0)}_3$ is exactly in the middle  between both.

\subsection{Auger line shapes including FA-PCI}
\label{ssec:fapci-lineshapes}
\begin{figure}
 \includegraphics[width=4.25cm]{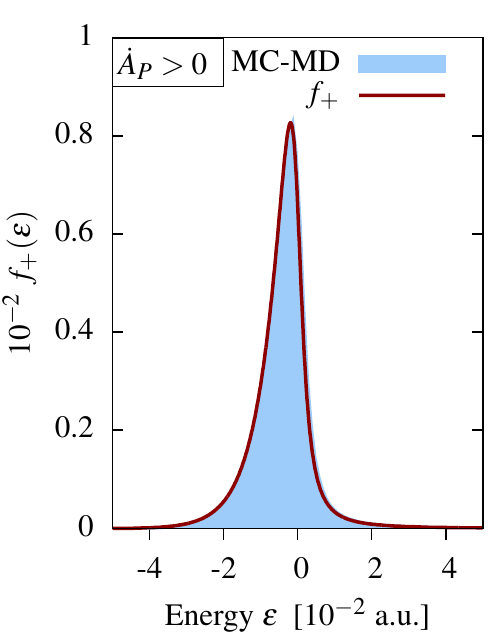}
 \includegraphics[width=4.25cm]{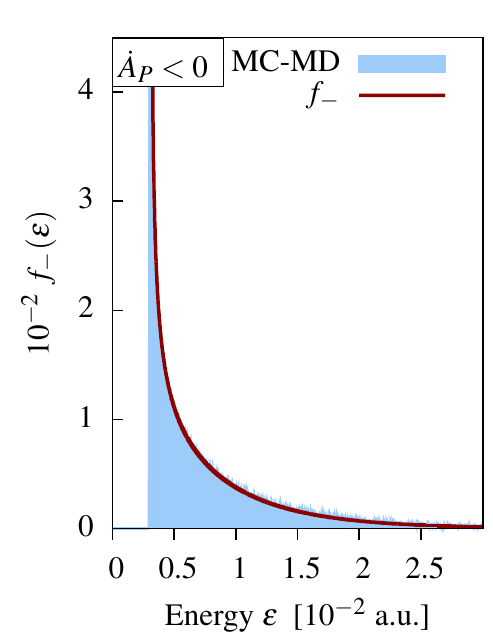}
 \caption{(color online) Auger lineshapes at zero transition of $A$ including streaking and PCI contributions. The analytical line shapes,
 Eqs.~\eqref{eq:lineshape-pci-plus} and~\eqref{eq:lineshape-pci-minus} [bold, red lines] are compared to MD simulations (filled area).
The case of $\dot{A}_P>0$ ($\dot{A}_P<0$) is shown in the left (right) panel. Parameters are the same as for Fig.~\ref{fig:pci-only}.}
 \label{fig:lineshapes-pci}
\end{figure}
Using the full mapping function \eqref{eq:epsilon-pci} gives a quadratic equation for $\tau_A$,
\begin{equation}
 \tau_A^2+\tau_A \left (\frac{p_r \epsilon^{\textup{PCI}}+\delta r^*p_{A_i}\dot{A}_P}{p_r p_{A_i} \dot{A}_P} \right ) + \frac{\delta r^* \epsilon^{\textup{PCI}} -1}{p_r p_{A_i}\dot{A_P}}=0 \;.
\end{equation}
For the inversion of Eq.~\eqref{eq:epsilon-pci}, we assume $p_r(\tau_A)\equiv p_r$, hence we neglect the additional implicit $\tau_A$ dependence, which
enters through the vector potential.
We define
\begin{equation}
 \epsilon_\pm \equiv \epsilon \pm \frac{\beta}{4} \delta r^*; \; \alpha\equiv \frac{1}{2 p_{A_i} \dot{A}_P}; \; \beta \equiv \frac{4 p_{A_i} \dot{A}_P}{p_r}
\end{equation}
and obtain
\begin{equation}
 \tau_{A\pm}=-\alpha \epsilon_+ \pm \alpha k_+ \; ,
 \label{eq:tauA-pci}
\end{equation}
with $k_{\pm}=\sqrt{\epsilon_-^2\pm|\beta|}$.
For $\dot{A}_P>0$, only the positive branch $\tau_{A+}$ can be realized ($\tau_A \geq 0$), which gives for the lineshape with $\Gamma_4=|\alpha|\Gamma_A$:
\begin{equation}
 f_+(\epsilon)=\Gamma_4 \frac{k_+-\epsilon_-}{k_+} e^{-\Gamma_4 (k_+ -\epsilon_+)} \;.
 \label{eq:lineshape-pci-plus}
\end{equation}

For $\dot{A}_P<0$, both solutions~\eqref{eq:tauA-pci}, $\tau_{A+}$ and $\tau_{A-}$, are possible. Thus,
the temporal distribution function \eqref{eq:time-distribution} is split into two parts,
 $\int_{0}^{\tau_{\textup{min}}}\; e^{-\Gamma_A\tau_A} \textup{d}\tau_A + \int_{\tau_{\textup{min}}}^\infty\; e^{-\Gamma_A\tau_A} \textup{d}\tau_A$, where
$\tau_{\textup{min}}$ separates both branches, $\tau_{A+}$ and $\tau_{A-}$, at $\epsilon(\tau_{\textup{min}})=\epsilon_0$.
The straightforward transformation of both integrals to energy gives for the joint energy distribution function
 \begin{equation}
 f_-(\epsilon)=2\Gamma_4 e^{-\Gamma_4 \epsilon_+} \left( \frac{\epsilon_-}{k_-} \cosh{\Gamma_4 k_-}  -\sinh{\Gamma_4 k_-} \right ) \;.
\label{eq:lineshape-pci-minus}
 \end{equation}
\begin{figure}
 \includegraphics[width=4.25cm]{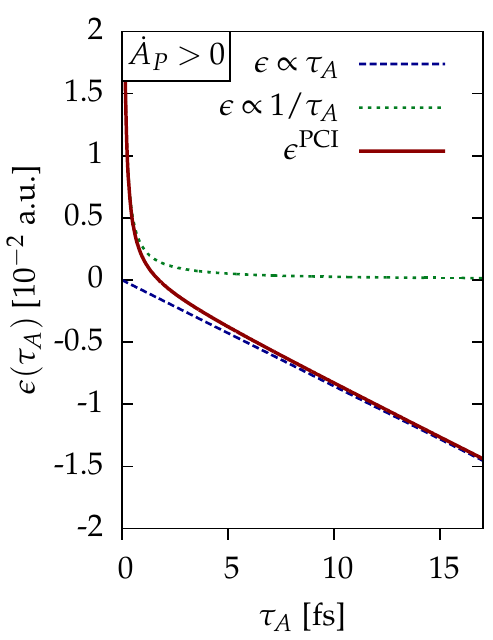}
 \includegraphics[width=4.25cm]{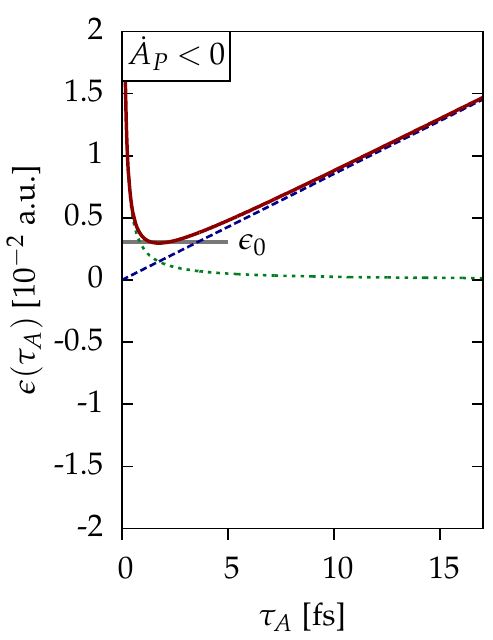}
 \caption{(color online) Time-to-energy mapping of the Auger electrons in the streaking field for $\dot{A}_P>0$ (left)
 and $\dot{A}_P<0$ (right). The mappings without PCI, Eq.~\eqref{eq:tauA-nopci-Azero} blue dashed lines, without streaking, 
Eq.~\eqref{eq:epsilon-pci-only} green dotted lines,  and including
PCI and streaking, Eq.~\eqref{eq:epsilon-pci} red solid lines, are shown. Different signs of $\dot{A}_P$ lead to a drastic change of
 the mapping: for $\dot{A}_P>0$ the whole energy space is accessible in a bijective way, whereas in contrast for $\dot{A}_P<0$ a forbidden
region below $\epsilon_0$ occurs and ``early'' and ``late'' Auger electrons are mapped to the same energy.
 }
 \label{fig:mechanism}
\end{figure}
The line shapes of Xe \emph{NOO} for both cases, $\dot{A}_P>0$ and $\dot{A}_P<0$, are given in Fig.~\ref{fig:lineshapes-pci}
 for $\delta r^*=1$. 
While for the former, the line is broadened by PCI, for the latter, the line is compressed
and completely different line shapes for subsequent zero transitions of $A$ with different sign of the slope are observed. 
As in the previous cases, perfect agreement with simulations based on numerical solutions
of Eqs.~(\ref{eq:eom_phase1}-\ref{eq:eom_phase3_ic}) by means of MC averaged MD simulations (in analogy to
Sec.~\ref{sec:semi-classics}, but without momentum averaging) is observed, and the linearization of $A$ has, 
in the considered regimes of pulse duration and ponderomotive potential,
 no significant influence on the streaked Auger spectra.

To explain the strikingly different shape of the Auger lines in Fig.~\ref{fig:lineshapes-pci}, the mapping functions from time to energy
are shown in Fig.~\ref{fig:mechanism} for the same set of parameters. The linear streaking part contributing to Eq.~\eqref{eq:epsilon-pci}
is plotted with blue dashed lines, the hyperbolic PCI term with green dotted lines and the sum of both results in the red solid lines.
By comparing $\dot{A}_P>0$ (left) and $\dot{A}_P<0$ (right), the cause for the different lineshapes becomes visible:
whereas for the former, both terms add up to a bijective mapping function spanning the whole energy axis from 
$-\infty$ to $\infty$, for the latter one a forbidden energy region for $\epsilon<\epsilon_0=2 \sqrt{p_A |\dot{A}_P|/p_r}$
occurs (gray line in Fig.~\ref{fig:mechanism}). This leads to a drastic compression
of the line (right panel in Fig.~\ref{fig:lineshapes-pci}), where Auger electrons released at two different time moments can be
 mapped into the same energy interval. This situation
is completely absent for $\dot{A}_P>0$ which leads to a broad distribution of Auger electrons, cf. left panel in Fig.~\ref{fig:lineshapes-pci}.
This effect is a direct consequence of the interplay between the hyperbolic PCI-induced chirp on the Auger electron energy,
 $\Delta E^{\textup{PCI}} \propto 1/r^* \propto 1/\tau_A$, and the linear ``chirp'' introduced by the streaking field,
 where the sign of the latter depends on the direction of the streaking field at the time of the core hole creation.
An experimental verification of this mechanism utilizing XUV pulses from FLASH and HHG exciting the Xe \emph{NOO} and Kr \emph{MNN} transitions
has been presented in Ref.~\cite{schuette_evidence_2011}. The comparison of the experimentally obtained Auger electron spectra
with the theoretical results calculated based on MC-MD simulations, as presented in Sec.~\ref{sec:semi-classics}, shows perfect agreement.

\subsection{Finite XUV pulse duration}
\label{ssec:finite-xuv}
\begin{figure}
 \includegraphics[width=4.25cm]{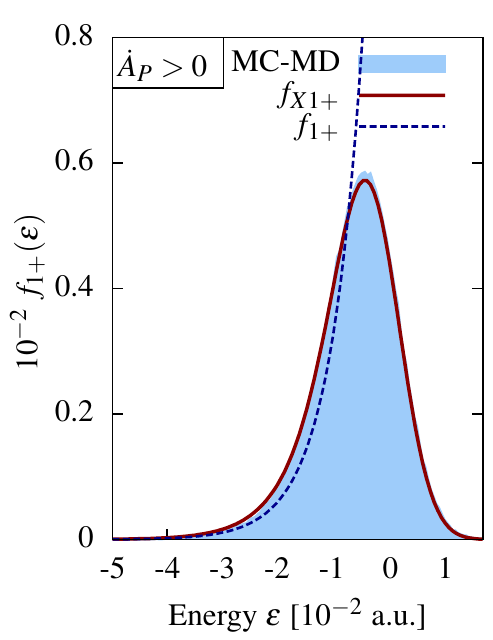}
 \includegraphics[width=4.25cm]{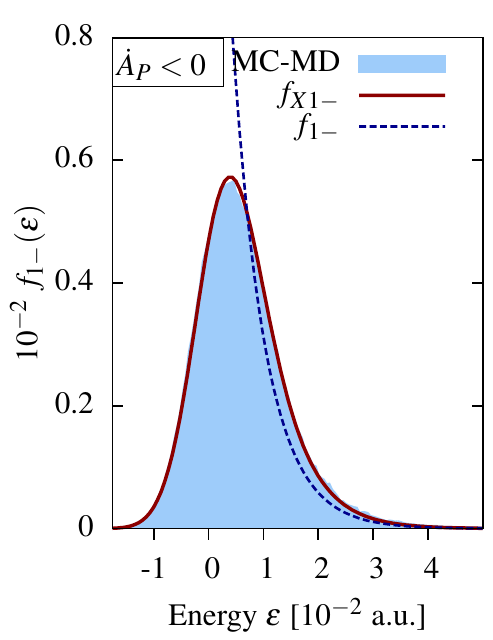}
 \caption{(color online) Auger lineshapes for zero transitions of $A$
 neglecting PCI contributions for a finite XUV pulse duration of $20\,$fs FWHM. All parameters refer to
THz streaking of the Xe \emph{NOO} transition.
The convoluted line shape, Eq.~\eqref{eq:lineshape-nopci-XUV} [bold red line],
 is plotted against MD simulations (blue area).
 The case of infinitesimal XUV pulse duration, Eq.~\eqref{eq:lineshape-nopci-azero}, is shown also for comparison (blue dashed lines).
}
 \label{fig:lineshapes-nopci-xuv}
\end{figure}
\begin{figure}
 \includegraphics[width=4.25cm]{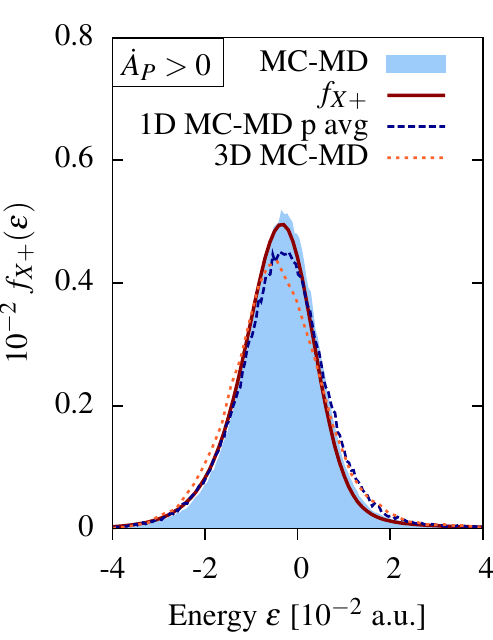}
 \includegraphics[width=4.25cm]{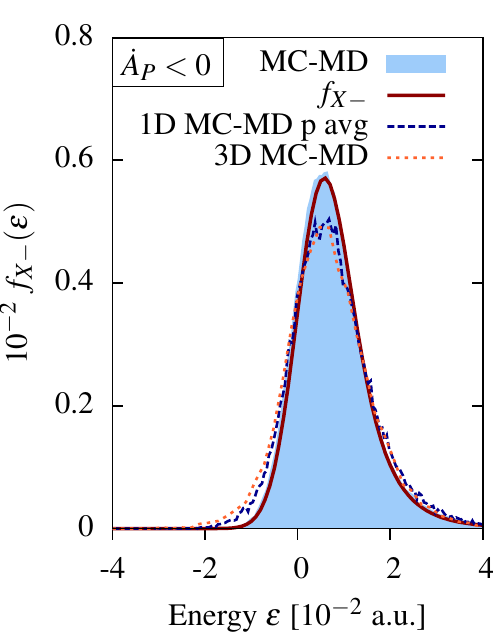}
 \caption{(color online) The same as Fig.~\ref{fig:lineshapes-nopci-xuv} but
for the case including PCI. The result of a numerical convolution, Eq.~\eqref{eq:lineshape-conv-pci},
red bold lines, is compared to MD simulations with sharp momenta (blue area) and
with proper sampling over initial momenta $p_{P_i}$ and $p_{A_i}$ (blue dashed lines).
Additionally, 3D MD simulations, cf. Sec.~\ref{sec:semi-classics}, are shown
(orange dotted lines).}
 \label{fig:lineshape-pci-XUV}
\end{figure}
Eqs.~\eqref{eq:lineshape-pci-plus} and \eqref{eq:lineshape-pci-minus} describe the shape of the Auger energy
distribution for infinitesimal pulse duration of the XUV excitation.
To account for finite pulse durations $\tau_X$, a similar transformation
of probability distributions from time to energy as for the case of the decay law \eqref{eq:time-distribution}
 needs to be performed.
The temporal distribution of a photoelectron excited by a Gaussian shaped pulse is described by
Eq.~\eqref{eq:xuv-time-distribution}.
At zero transitions of the vector potential, utilizing the same linearization of $A$ as was used for Eq.~\eqref{eq:epsilon},
 we obtain for the $\tau_i$-dependent energy shift (see Fig.~\ref{fig:times} for notations) 
\begin{equation}
 \tilde{\epsilon}(\tau_i)=-p_i \dot{A}(t_X)\tau_i + \mathcal{O}(A^2,\tau_i^2)\;.
\end{equation}
Using this mapping function, a streaked energy spectrum due to the finite XUV pulse duration can be calculated:
\begin{eqnarray}
 f_X(\tilde{\epsilon})=\frac{1}{\sqrt{\pi} \Gamma_{X1}} \exp \left (- \frac{\tilde{\epsilon}^2}{\Gamma_{X1}^2} \right ) \; 
 \label{eq:xuv-distribution}
\end{eqnarray}
with the normalization condition
\begin{eqnarray}
   \int_{-\infty}^{\infty} \textup{d}\tilde{\epsilon} \; f_{X}(\tilde{\epsilon})=1,\;  \textup{and}\;\;\;\Gamma_{X1} = p_{A_i} \dot{A}(t_X) \tau_X \;.
\end{eqnarray}
At each photoelectron ``birth'' time $t_{P_i}$, the Auger ``clock'' starts, and with that the 
energy mapping of the temporal distribution of Auger electrons. Therefore, the final streaked Auger energy 
distribution is given by the convolution
\begin{equation}
 f_{XA}(\epsilon)=\int_{-\infty}^{\infty} \textup{d}\tilde{\epsilon} \; f_{X} (\tilde{\epsilon}) f_A (\epsilon-\tilde{\epsilon}) \;.
 \label{eq:lineshape-convolution}
\end{equation}
If the PCI contribution is neglected, i.~e.~$f_A(\epsilon)=f_{1\pm}(\epsilon)$, cf. Eq.~\eqref{eq:lineshape-nopci-azero},
the integration in Eq.~\eqref{eq:lineshape-convolution} can be carried out analytically and gives for the line shape
\begin{equation}
 f_{X1\pm}(\epsilon) = \frac{\Gamma_1}{2} e^{\frac{1}{4} \Gamma_{X1}^2 \Gamma_1^2} e^{\pm \Gamma_1 \epsilon} \textup{erfc}\left( \frac{\Gamma_{X1} \Gamma_1}{2} \pm \frac{\epsilon}{\Gamma_{X1}}\right )\; .
 \label{eq:lineshape-nopci-XUV}
\end{equation}
The lineshape for two subsequent zero transitions of $A$ for an XUV pulse duration of $20\,$fs FWHM is shown in 
Fig.~\ref{fig:lineshapes-nopci-xuv} (bold red line). The finite excitation time interval of the core hole broadens the pure Auger decay line (blue 
dashed line). Despite the rather long pulse duration compared to the core hole lifetime, the exponential decay of
the case of infinitesimal excitation duration is still imprinted on the convoluted line. Corresponding 
MD data  (blue area, according to method (i) below) is
in perfect agreement with Eq.~\eqref{eq:lineshape-nopci-XUV}.

Considering the PCI-distorted line shapes, $f_A(\epsilon)=f_{\pm}(\epsilon)$, the integral 
\begin{equation}
 f_{X\pm}(\epsilon) = \int_{-\infty}^{\infty} \textup{d} \tilde{\epsilon} \; f_X(\tilde{\epsilon}) f_{\pm}(\epsilon-\tilde{\epsilon}) 
 \label{eq:lineshape-conv-pci}
\end{equation}
cannot be solved analytically.
The result of a numerical integration of Eq.~\eqref{eq:lineshape-conv-pci} 
for a $20\,$fs FWHM XUV pulse is given in Fig.~\ref{fig:lineshape-pci-XUV} for positive and negative slope of $A(t)$.
Although the asymmetry with respect to $\dot{A}_P>0$ and $\dot{A}_P<0$ is less pronounced 
than for the case of infinitesimal XUV excitation duration shown in Fig.~\ref{fig:lineshapes-pci},
still a difference between ascending and descending slope of $A$ is visible indicating the chirp in Auger emission.
To verify the accuracy of the analytical result, additionally three different sets of simulations are shown:
(i) numerical solutions according to Eqs.~(\ref{eq:eom_phase1}-\ref{eq:eom_phase3_ic}) [(blue) filled area], 
(ii) similar to (i) but with proper averaging over the
initial momenta $p_{A_i}$ and $p_{p_i}$ [blue dashed lines] and (iii) 
3D MD simulations according to the scheme 
in Sec.~\ref{sec:semi-classics}, also including angular distributions [orange dotted lines].
(i) resembles the assumptions of the analytical model, except for the linearization of $A$, and shows 
perfect agreement with Eq.~\eqref{eq:lineshape-conv-pci}.
 Solutions according to (ii) and (iii) show a substantial broadening of the Auger line. For (ii) 
this results from the natural Auger line width and the bandwidth of the XUV pulse, and for (iii) in addition 
from the angle integration.
This broadening occurs in a similar way for rising and falling flank of $A$ and does not affect
the observed asymmetry attributed to the PCI-induced chirp.
Therefore, the analytical line shape model \eqref{eq:lineshape-conv-pci} catches the important features of
FA-PCI. It can be evaluated numerically for a large set of parameters due to its simple convolution structure.
Thus, Eq.~\eqref{eq:lineshape-conv-pci} is well-suited for the detailed investigation of the properties of
FA-PCI and its dependence on the streaking conditions and XUV parameters.

\subsection{Photoelectron distributions}
\label{ssec:pe-distribution}
\begin{figure}
 \includegraphics[width=8.5cm]{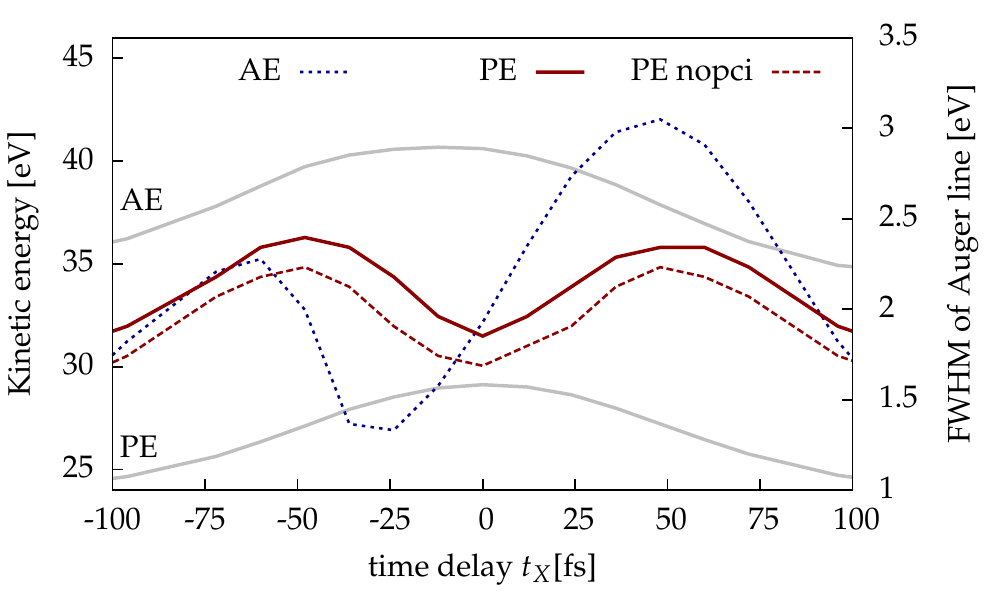}
 \caption{(color online) Energy (left axis) and FWHM (right axis) of the Auger electron (dotted)
 and the photoelectron distribution (PCI: solid, without PCI: dashed). 
Shown is the case of Xe \emph{NOO} scaled by $\gamma=10$ obtained within 
TDSE simulations. Parameters are the same as for Fig.~\ref{fig:tdse-duration}.}
 \label{fig:width-pe}
\end{figure}

In the previous sections, we identified the physical mechanism for the observed time-dependent chirp on the
Auger electron's energy: a direct connection between the time instant of decay and an associated (unique) energy
shift mediated through post-collision interaction. At this point, a remark on the consequences for the 
corresponding photoelectron distribution is appropriate.
As a matter of fact, the kinetic energy of the photoelectron is affected in a similar way as it is for the Auger electron,
but, by reason of energy conservation, with an opposite sign. Thus, the photoelectron is slowed down due to PCI
by the same amount of energy $\Delta E^{\textup{PCI}}$ the Auger electron has gained.

However, this does not result, as one might guess, in a chirp on the photoelectron energy 
distribution with different sign, as already pointed out in Sec.~\ref{ssec:tdse-pe}.
 The results of TDSE simulations, carried out as described in Sec.~\ref{sec:tdse}, 
are shown in Fig.~\ref{fig:width-pe}.
The FWHM for the photoelectron line is depicted for the case including PCI (solid line) and neglecting PCI (dashed line)
for a full set of time delays between pump and probe pulse. For both cases, no asymmetry with respect to the rising and the falling flank
of the vector potential is observed. Only a broadening of the line for the PCI-included case is present, resulting in 
an equidistant upward shift of the PCI curve in comparison to the case neglecting PCI, and therefore, no chirp on the photoelectron
energy can be identified.
The prominent asymmetry in the FWHM for the Auger electron is shown for comparison with dotted lines in Fig.~\ref{fig:width-pe}.

Returning to Eq.~\eqref{eq:epsilon-pci}, the mapping from the time moment of Auger decay, $\tau_A$, to energy, the origin
for the absence of a chirp becomes clear: while for Auger electrons, each decay time corresponds to a certain amount
of energy transfer, for photoelectrons no such connection can be found. For each time moment of photoemission, every Auger decay 
time $\tau_A$ is possible, and with that any arbitrary energy transfer due to PCI. Thus, only a PCI broadening of the line 
is expected, which is the same for every release time of the photoelectron and explains the observed delay dependence
of the FWHM of the photoelectron spectrum in Fig.~\ref{fig:width-pe}.

\section{Use of FA-PCI for pulse characterization}
\label{sec:params}

\begin{figure}
 \includegraphics[width=4.25cm]{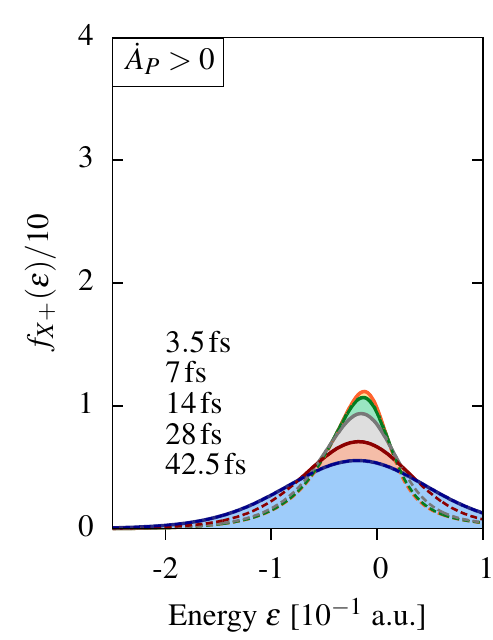}
 \includegraphics[width=4.25cm]{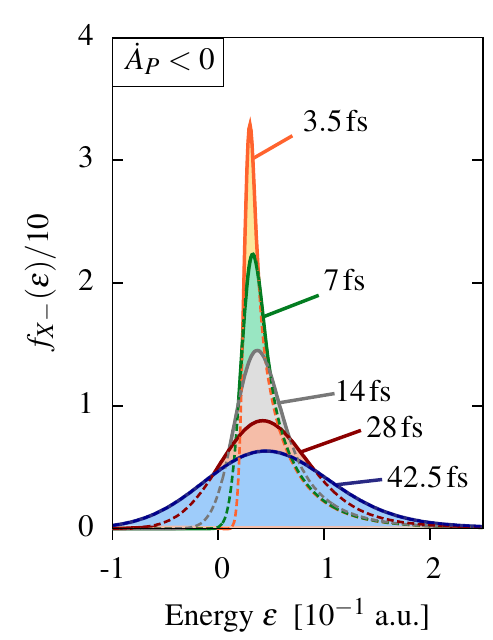}
 \caption{(color online) Auger line shape of the Kr \emph{MNN} transition in a $3.3\,$THz
streaking field at fixed ponderomotive potential of $100\,$meV  for different XUV pulse durations
with a photon energy of $97\,$eV. Shown are solutions of the analytical model,
 Eq.~\eqref{eq:lineshape-conv-pci}.}
 \label{fig:lineshape-pci-taudep}
\end{figure}
\begin{figure}
 \includegraphics[width=4.25cm]{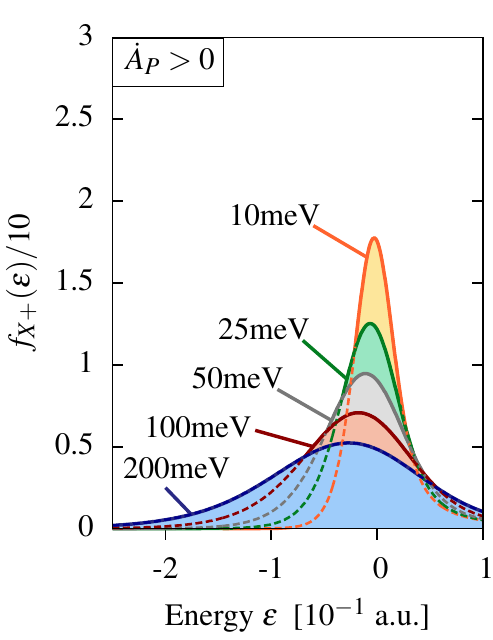}
 \includegraphics[width=4.25cm]{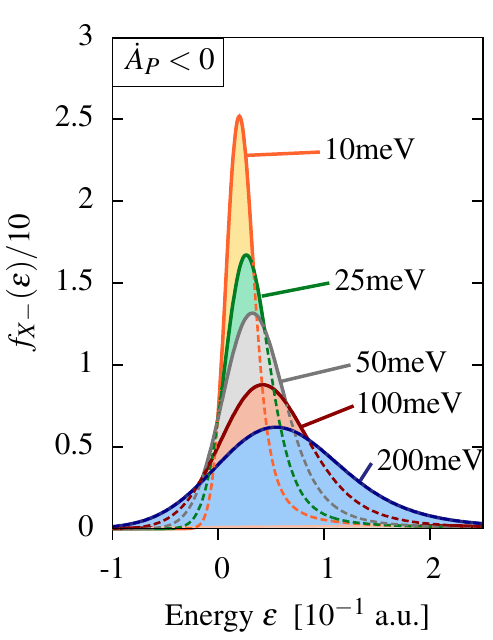}
 \caption{(color online) The same as Fig.~\ref{fig:lineshape-pci-taudep} but for different ponderomotive
potential of the streaking field at fixed XUV pulse duration of $28\,$fs.}
 \label{fig:lineshape-pci-updep}
\end{figure}

In this last part we ask the question, whether the PCI-induced asymmetry with respect to the direction
of the slope of the vector potential may help to improve the pulse characterization capabilities of the streak
camera.
In Sec.~\ref{sec:tdse} we already commented on the influence of the ponderomotive potential and the XUV
pump pulse duration on the observed asymmetry. This raises the question whether this dependence, especially
on the pulse duration $\tau_X$, can be used as a sensitive parameter to recover the pulse durations in experiments.
 Due to the larger effect of PCI, we choose Kr \emph{MNN} decay
at a photon energy of $97\;$eV in the following.

The line shape for different XUV pulse durations at the zero transitions of the vector potential
with $U_p=100\,$meV is shown in Fig.~\ref{fig:lineshape-pci-taudep}, 
calculated according to Eq.~\eqref{eq:lineshape-conv-pci}.
The left panel shows the broadened line $f_X^+$ for $\dot{A}>0$ and the right panel the corresponding 
compressed line $f_X^-$ for $\dot{A}<0$.
For the shortest pulse durations ($3.5$ and $7\,$fs) with $\tau_X \lesssim \Gamma_A^{-1}$
 the largest asymmetry is observed, whereas for long pulses ($42\,$fs) no
clear distinction between $\dot{A}>0$ and $\dot{A}<0$ is possible. The largest impact on this asymmetry
has the compressed line, which is sharp in the case of very short pump pulses, cf. Fig.~\ref{fig:lineshapes-pci}, right panel.
For increasing $\tau_X$, the convolution with the Gaussian shaped time distribution of the XUV excitation smoothens
(broadens) this line, until the streaking-induced broadening predominates. This result agrees qualitatively 
with the TDSE simulations presented above in Sec.~\ref{ssec:tdse-results}, cf. Fig.~\ref{fig:tdse-duration}.

A similar behavior is observed upon change of the ponderomotive potential of the streaking pulse at a fixed
pump pulse duration of $28\,$fs, see Fig.~\ref{fig:lineshape-pci-updep}. Here, the largest 
asymmetry is observed for small $U_p$, whereas the asymmetry gradually decreases upon increase of $U_p$.
This effect can be as well explained by a domination of the streaking contribution for 
large $U_p$, where the broadening of the line due to the larger momentum transfer from the streaking field
exceeds the PCI contribution. Again, a similar picture arises in TDSE calculations, cf. Fig.~\ref{fig:tdse-up}.

In a further step, we want to evaluate the asymmetry in more detail.
To this end, we introduce a classification parameter $\xi$, defined as
\begin{equation}
 \xi=\frac{|w^--w^+|}{w^-+w^+}\;,
\label{eq:asym-param}
\end{equation}
where $w^\pm$ is the FWHM of the Auger line at $\dot{A}>0$ and $\dot{A}<0$, respectively.
This parameter  $\xi \in [0,1]$ describes the relative asymmetry and is zero for vanishing 
asymmetry and attains finite 
positive values smaller than one in any other case.

\subsection{Effect of XUV pump pulse duration}
\begin{figure}
 \includegraphics[width=0.48\textwidth]{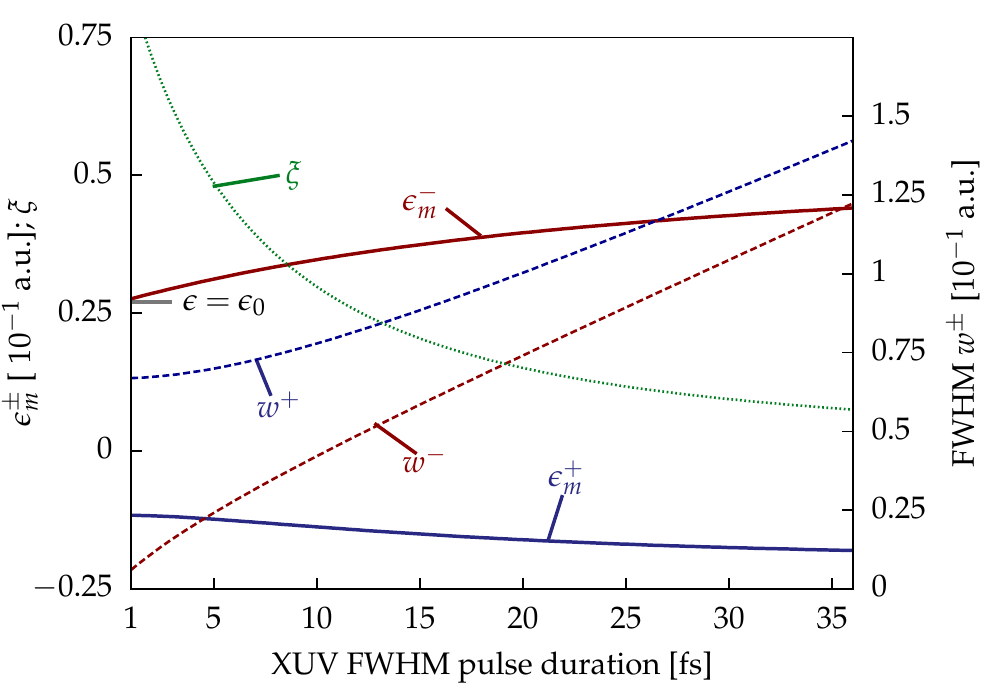}
 \caption{(color online) Position of the Auger line's maximum $\epsilon_m^\pm$,
the corresponding width $w^\pm$ and the asymmetry parameter $\xi$, cf.
Eq.~\eqref{eq:asym-param}, vs. XUV pump pulse duration.
Data is for Kr \emph{MNN} decay at a ponderomotive potential of $100\,$meV of the streaking pulse.
The forbidden region for $\dot{A}_p<0$ at infinitesimal short XUV pulse duration is marked by $\epsilon_0$.
The superscript ''+``(''-``) refers to $\dot{A}_p>0$ ($\dot{A}_p<0$). }
 \label{fig:asymmetry-model-tau}
\end{figure}
A key property for successful pulse characterization is the existence of observables
sensitive to the XUV pulse duration.
In Fig.~\ref{fig:asymmetry-model-tau} several properties of the Auger line shapes are plotted
depending on the XUV pulse duration at fixed $U_p=100\,$meV:
the position of the maximum of the line, $\epsilon_m^{\pm}$, the FWHM of the line, $w^\pm$,
 and the corresponding asymmetry parameter $\xi$.
Let us first consider the energetic position of the line. 
For $\dot{A}<0$, $\epsilon_m^-$  is positive,
stemming directly from the positive tail of the decay law. For small pulse durations, the forbidden
region $\epsilon < \epsilon_0$, discussed in Sec.~\ref{ssec:fapci-lineshapes}, is observed with the 
sharp onset of the line at $\epsilon_0$. With increasing 
pulse duration, the position of the maximum shifts towards higher energies, a clear consequence of the convolution
with the Gaussian shaped temporal distribution of the XUV pulse. This shift saturates for large 
pulse durations, $\tau_X \gg \Gamma_A^{-1}$, due to the broad convolution function.
For $\dot{A}>0$ and $\epsilon_m^+$, a similar trend is observed for $\epsilon_m^+$, but with decreasing energy of the maximum.
Due to the PCI-induced broadening of the line, the effect is here less pronounced than for the compressed line.

The width of the line, $w^\pm$, shows the typical strong influence on the XUV pump pulse duration: the 
larger $\tau_X$, the larger $w^\pm$. This phenomenon is the basic principle of the streak camera
utilized for the estimation  of pulse lengths. 
Here, the observed asymmetry due to the PCI-induced chirp on the Auger electron's energy
manifests itself in two branches for the width: one for $\dot{A}>0$ labeled by $w^+$ and one for
$\dot{A}<0$ indicated by $w^-$, cf. dashed lines in Fig.~\ref{fig:asymmetry-model-tau}. 
For conventional chirp-free situations
both branches coincide (no asymmetry with respect to $\dot{A}>0$ and $\dot{A}<0$).

Since with two opposing detectors, both situations $\dot{A}>0$ and $\dot{A}<0$ can be recorded
simultaneously, also the asymmetry parameter $\xi$ carries valuable information about the single-shot pulse
properties. Its pulse-duration dependence is plotted by the (green) dotted line:
for short pulses a large asymmetry of about $0.75$ is observed with a 
rapid decrease down to about $0.1$ at pulse durations of $20\,$fs. The largest variation is found for pulses
below $10\,$fs where $\tau_X$ is comparable to the Auger decay time $\Gamma_A^{-1}$.
Thus from measuring the width of the Auger lines at opposite slopes of the streaking vector potential 
simultaneously, a reconstruction of the pulse duration is possible, even if
PCI effects are present.

\subsection{Ponderomotive potential of streaking pulse}
\begin{figure}
  \includegraphics[width=0.48\textwidth]{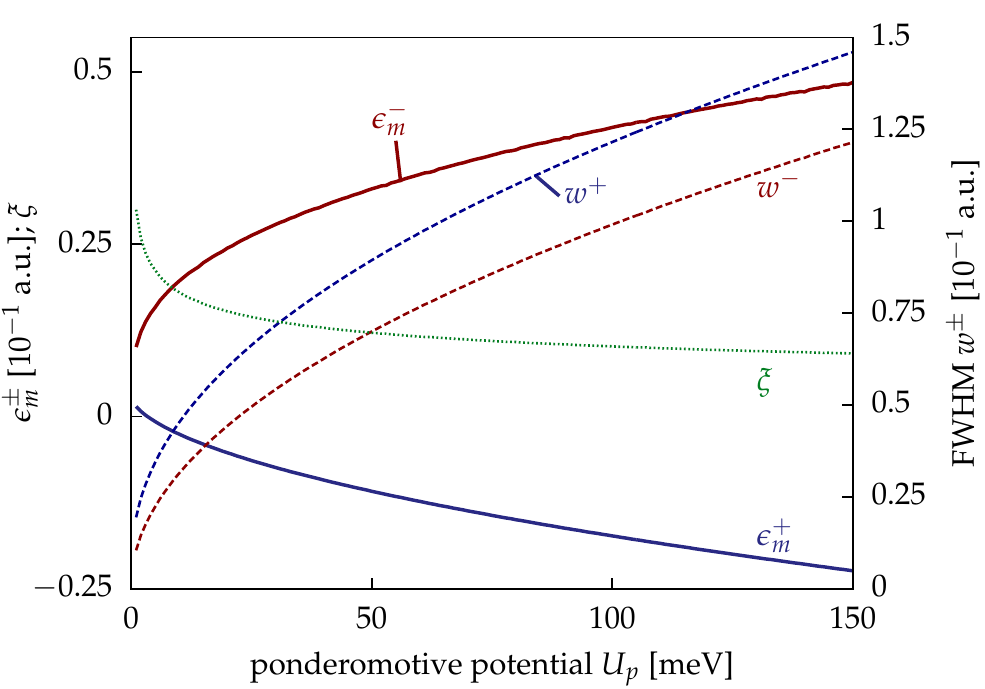}
  \caption{(color online) The same as Fig.~\ref{fig:asymmetry-model-tau} but for varied
ponderomotive potential $U_p$ of the streaking field at a fixed pump pulse duration of $28\;$fs.}
  \label{fig:asymmetry-model-up}
\end{figure}

\begin{figure}
  \includegraphics[width=0.48\textwidth]{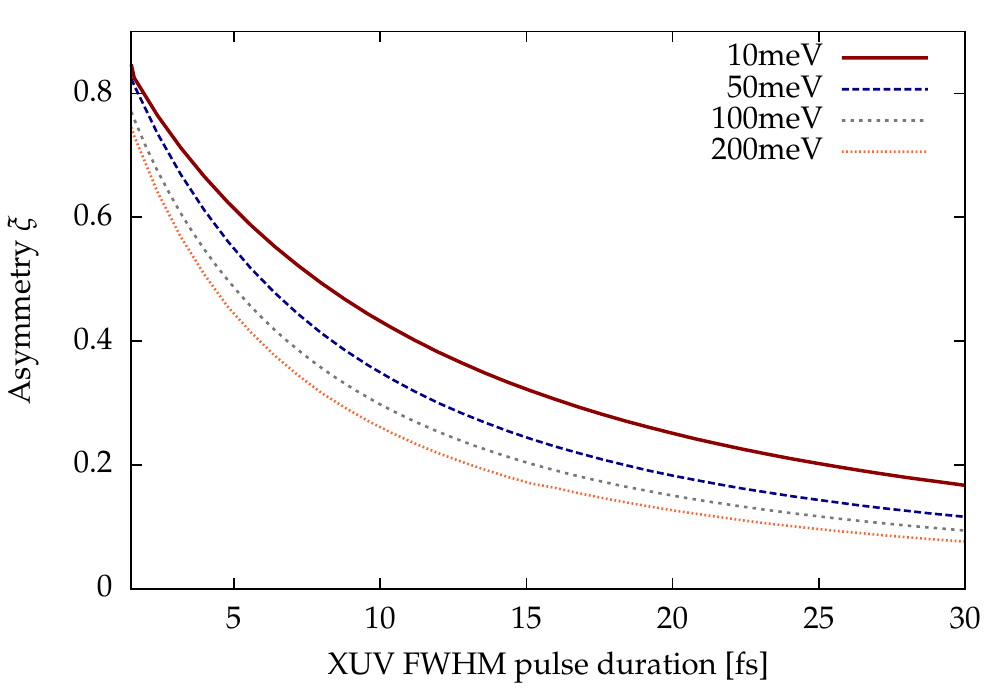}
  \caption{(color online) Asymmetry parameter $\xi$ for different ponderomotive potentials of the streaking pulse as a function of the XUV pulse duration.
Parameters are the same as for Fig.~\ref{fig:asymmetry-model-tau}.}
\label{fig:asymmetry-model-tau-up}
\end{figure}
An important question is the dependence of the asymmetry on the streaking conditions and, in 
particular, the ponderomotive potential $U_p$ of the streaking field. This
parameter is, in principle, easily tunable in experiments, either through the frequency $\omega_L$
(limited by the pulse duration $\tau_X \ll \omega_L^{-1}$) or the intensity $E_L^2$.
In Fig.~\ref{fig:asymmetry-model-up} the position of the maximum, the FWHM and the asymmetry parameter
for a scan of $U_p$, by variation of $E_L$, at a fixed XUV pulse duration of $28\,$fs are shown.
For vanishing $U_p$, no streaking occurs, which gives vanishing energy shifts $\epsilon_m^\pm$ 
and widths $w^\pm$. [Note: no natural line widths are included in Eq.~\eqref{eq:lineshape-conv-pci}].

For increasing $U_p$, the position of the maximum of the line shifts towards higher energies
 for the case of $\dot{A}<0$ ($\epsilon_m^-$) and towards smaller energies for $\dot{A}>0$ ($\epsilon_m^+$). 
This effect is similar to the behavior observed
upon variation of $\tau_X$. Also, as in the previous case, two branches of the width
can be identified, originating from the asymmetry (dashed lines in Fig.~\ref{fig:asymmetry-model-up}).
For larger $U_p$, the momentum transfer from the probing field to the electron increases and, with that,
the streaking, resulting in broader lines. Again for the case of chirp-free XUV excitation without 
PCI effects the ''+`` and the ''-`` branches would coincide.

Additionally, the asymmetry parameter $\xi$ is plotted in Fig.~\ref{fig:asymmetry-model-up}.
Starting from a rather high value at very small $U_p$, it exhibits
a rapid drop when $U_p$ increases to about $10\,$meV, followed by a slow convergence with only a weak dependence on $U_p$
over a wide range of approximately $100\,$meV.

 Fig.~\ref{fig:asymmetry-model-tau-up} summarizes
the possibilities for pulse characterization using FA-PCI. The asymmetry parameter $\xi$ is shown
as a function of $\tau_X$ for different $U_p$ of the streaking pulse. For all considered values 
of $U_p$ a monotonic behavior with large asymmetry for small pulse durations and vice versa
is observed. The larger $U_p$ the faster is the drop of $\xi$ at small pulse durations and, with the
corresponding strong variation with $\tau_X$, a high sensitivity of $\xi$ in the range of
pulse durations below $10$ to $20\,$fs occurs. Thus increasing $U_p$ allows to extend the region of sensitivity
to slightly larger XUV pulse durations.
In conclusion, measuring $\xi$  at a given ponderomotive potential of the streaking field for  $\dot{A}>0$ and $\dot{A}<0$ 
simultaneously utilizing opposite detectors, an estimation of 
the XUV pulse duration based on (time-resolved) Auger electron spectroscopy is possible. The highest sensitivity is reached
for XUV pulse durations below $10\,$fs with a rather strong variation of the measured parameter $\xi$ by a factor of four.

\section{Conclusions and Outlook}
\label{sec:conclusions}
In this paper, we gave a detailed theoretical explanation of the experimental observations in \cite{schuette_evidence_2011},
which show evidence of an energetic chirp in Auger emission.
Based on solutions of the TDSE we could reproduce the chirp for model systems and explain its
origin by post-collision interaction. This formed the basis for classical modeling of the photoelectron and the Auger electron
in the continuum, including all electron-electron and electron-ion interactions.
Using Monte-Carlo averaged Molecular Dynamics simulations for the electrons, we were able to verify this chirp. The 
quantitative comparison with experiments including detector resolutions and acceptance
geometries using the Xe \emph{NOO} and the Kr \emph{MNN}
transitions presented in \cite{schuette_evidence_2011} shows perfect agreement between our approach 
and the light-field driven streak camera in the considered range of parameters.
For deeper insight and to obtain a more flexible tool, we derived a classical, 
analytical line shape model for the Auger electron
that fully includes the XUV pulse shape, streaking and PCI effects and thus captures all important properties. 
We further showed, how our results
may be used as a tool for estimating the length of XUV pump pulses if PCI effects are involved.

In the present work, we focused on the Auger electrons, which was motivated by currently available
experiments. The corresponding photoelectron distribution was briefly discussed, and we explained why no 
energetic chirp is present there. A detailed analysis will be part of a future work.
Worthwhile considerations include the influence of additional XUV pulse parameters such as chirp and 
substructures, e.~g.~spikes as present in the case of free electron laser sources. 
Additionally, it will be advantageous to extend our purely classical model for the FA-PCI line shape to account for quantum
effects in order to describe interference and spin effects.

Finally, it would be very interesting to investigate in experiments with either Kr \emph{MNN} or Xe \emph{NOO}
the behavior when the XUV photon energy is increased.
If the proposed FA-PCI scenario is correct, then the chirp of the Auger spectra should vanish when
the photoelectron energy starts to exceed the Auger electron energy.

\acknowledgments{We thank B. Sch\"utte, U. Fr\"uhling and M. Drescher
for many interesting discussions of their experiments. We are grateful to N. Kabachnik 
for bringing to our attention the early work on PCI, in particular Ref.~\cite{ogurtsov_auger_1983}.
This work has been supported by the BMBF-Verbund ''FLASH'' and 
grant shp0006 for computer time at HLRN.}

\begin{thebibliography}{10}%
\makeatletter
\providecommand \@ifxundefined [1]{%
 \ifx #1\undefined \expandafter \@firstoftwo
 \else \expandafter \@secondoftwo
\fi
}%
\providecommand \@ifnum [1]{%
 \ifnum #1\expandafter \@firstoftwo
 \else \expandafter \@secondoftwo
\fi
}%
\providecommand \enquote [1]{``#1''}%
\providecommand \bibnamefont  [1]{#1}%
\providecommand \bibfnamefont [1]{#1}%
\providecommand \citenamefont [1]{#1}%
\providecommand\href[0]{\@sanitize\@href}%
\providecommand\@href[1]{\endgroup\@@startlink{#1}\endgroup\@@href}%
\providecommand\@@href[1]{#1\@@endlink}%
\providecommand \@sanitize [0]{\begingroup\catcode`\&12\catcode`\#12\relax}%
\@ifxundefined \pdfoutput {\@firstoftwo}{%
 \@ifnum{\z@=\pdfoutput}{\@firstoftwo}{\@secondoftwo}%
}{%
 \providecommand\@@startlink[1]{\leavevmode}%
 \providecommand\@@endlink[0]{}%
}{%
 \providecommand\@@startlink[1]{%
  \leavevmode
  \pdfstartlink
   attr{/Border[0 0 1 ]/H/I/C[0 1 1]}%
   user{/Subtype/Link/A<</Type/Action/S/URI/URI(#1)>>}%
  \relax
 }%
 \providecommand\@@endlink[0]{\pdfendlink}%
}%
\providecommand \url  [0]{\begingroup\@sanitize \@url }%
\providecommand \@url [1]{\endgroup\@href {#1}{\urlprefix}}%
\providecommand \urlprefix [0]{URL }%
\providecommand \Eprint[0]{\href }%
\@ifxundefined \urlstyle {%
  \providecommand \doi [1]{doi:\discretionary{}{}{}#1}%
}{%
  \providecommand \doi [0]{doi:\discretionary{}{}{}\begingroup
  \urlstyle{rm}\Url }%
}%
\providecommand \doibase [0]{http://dx.doi.org/}%
\providecommand \Doi[1]{\href{\doibase#1}}%
\providecommand \bibAnnote [3]{%
  \BibitemShut{#1}%
  \begin{quotation}\noindent
    \textsc{Key:}\ #2\\\textsc{Annotation:}\ #3%
  \end{quotation}%
}%
\providecommand \bibAnnoteFile [2]{%
  \IfFileExists{#2}{\bibAnnote {#1} {#2} {\input{#2}}}{}%
}%
\providecommand \typeout [0]{\immediate \write \m@ne }%
\providecommand \selectlanguage [0]{\@gobble}%
\providecommand \bibinfo [0]{\@secondoftwo}%
\providecommand \bibfield [0]{\@secondoftwo}%
\providecommand \translation [1]{[#1]}%
\providecommand \BibitemOpen[0]{}%
\providecommand \bibitemStop [0]{}%
\providecommand \bibitemNoStop [0]{.\EOS\space}%
\providecommand \EOS [0]{\spacefactor3000\relax}%
\providecommand \BibitemShut [1]{\csname bibitem#1\endcsname}%
\bibitem{brabec_intense_2000}%
  \BibitemOpen
  \bibfield{author}{%
  \bibinfo {author} {\bibfnamefont{T.}~\bibnamefont{Brabec}}\ and\ \bibinfo
  {author} {\bibfnamefont{F.}~\bibnamefont{Krausz}},\ }%
  \bibfield{journal}{%
  \bibinfo {journal} {Rev. Mod. Phys.}\ }%
  \textbf{\bibinfo {volume} {72}},\ \bibinfo {pages} {545} (\bibinfo {year}
  {2000})%
  \bibAnnoteFile{NoStop}{brabec_intense_2000}%
\bibitem{agostini_physics_2004}%
  \BibitemOpen
  \bibfield{author}{%
  \bibinfo {author} {\bibfnamefont{P.}~\bibnamefont{Agostini}}\ and\ \bibinfo
  {author} {\bibfnamefont{L.~F.}\ \bibnamefont{{DiMauro}}},\ }%
  \bibfield{journal}{%
  \bibinfo {journal} {Rep. Prog. Phys.}\ }%
  \textbf{\bibinfo {volume} {67}},\ \bibinfo {pages} {813} (\bibinfo {year}
  {2004})%
  \bibAnnoteFile{NoStop}{agostini_physics_2004}%
\bibitem{scrinzi_attosecond_2006}%
  \BibitemOpen
  \bibfield{author}{%
  \bibinfo {author} {\bibfnamefont{A.}~\bibnamefont{Scrinzi}}, \bibinfo
  {author} {\bibfnamefont{M.~Y.}\ \bibnamefont{Ivanov}}, \bibinfo {author}
  {\bibfnamefont{R.}~\bibnamefont{Kienberger}},\ and\ \bibinfo {author}
  {\bibfnamefont{D.~M.}\ \bibnamefont{Villeneuve}},\ }%
  \bibfield{journal}{%
  \bibinfo {journal} {J. Phys. B: At. Mol. Opt. Phys.}\ }%
  \textbf{\bibinfo {volume} {39}},\ \bibinfo {pages} {R1} (\bibinfo {year}
  {2006})%
  \bibAnnoteFile{NoStop}{scrinzi_attosecond_2006}%
\bibitem{krausz_attosecond_2009}%
  \BibitemOpen
  \bibfield{author}{%
  \bibinfo {author} {\bibfnamefont{F.}~\bibnamefont{Krausz}}\ and\ \bibinfo
  {author} {\bibfnamefont{M.}~\bibnamefont{Ivanov}},\ }%
  \bibfield{journal}{%
  \bibinfo {journal} {Rev. Mod. Phys.}\ }%
  \textbf{\bibinfo {volume} {81}},\ \bibinfo {pages} {163} (\bibinfo {year}
  {2009})%
  \bibAnnoteFile{NoStop}{krausz_attosecond_2009}%
\bibitem{goulielmakis_direct_2004}%
  \BibitemOpen
  \bibfield{author}{%
  \bibinfo {author} {\bibfnamefont{E.}~\bibnamefont{Goulielmakis}}, \bibinfo
  {author} {\bibfnamefont{M.}~\bibnamefont{Uiberacker}}, \bibinfo {author}
  {\bibfnamefont{R.}~\bibnamefont{Kienberger}}, \bibinfo {author}
  {\bibfnamefont{A.}~\bibnamefont{Baltuska}}, \bibinfo {author}
  {\bibfnamefont{V.}~\bibnamefont{Yakovlev}}, \bibinfo {author}
  {\bibfnamefont{A.}~\bibnamefont{Scrinzi}}, \bibinfo {author}
  {\bibfnamefont{T.}~\bibnamefont{Westerwalbesloh}}, \bibinfo {author}
  {\bibfnamefont{U.}~\bibnamefont{Kleineberg}}, \bibinfo {author}
  {\bibfnamefont{U.}~\bibnamefont{Heinzmann}}, \bibinfo {author}
  {\bibfnamefont{M.}~\bibnamefont{Drescher}},\ and\ \bibinfo {author}
  {\bibfnamefont{F.}~\bibnamefont{Krausz}},\ }%
  \bibfield{journal}{%
  \bibinfo {journal} {Science}\ }%
  \textbf{\bibinfo {volume} {305}},\ \bibinfo {pages} {1267} (\bibinfo {year}
  {2004})%
  \bibAnnoteFile{NoStop}{goulielmakis_direct_2004}%
\bibitem{uiberacker_attosecond_2007}%
  \BibitemOpen
  \bibfield{author}{%
  \bibinfo {author} {\bibfnamefont{M.}~\bibnamefont{Uiberacker}}, \bibinfo
  {author} {\bibfnamefont{T.}~\bibnamefont{Uphues}}, \bibinfo {author}
  {\bibfnamefont{M.}~\bibnamefont{Schultze}}, \bibinfo {author}
  {\bibfnamefont{A.~J.}\ \bibnamefont{Verhoef}}, \bibinfo {author}
  {\bibfnamefont{V.}~\bibnamefont{Yakovlev}}, \bibinfo {author}
  {\bibfnamefont{M.~F.}\ \bibnamefont{Kling}}, \bibinfo {author}
  {\bibfnamefont{J.}~\bibnamefont{Rauschenberger}}, \bibinfo {author}
  {\bibfnamefont{N.~M.}\ \bibnamefont{Kabachnik}}, \bibinfo {author}
  {\bibfnamefont{H.}~\bibnamefont{Schr\"oder}}, \bibinfo {author}
  {\bibfnamefont{M.}~\bibnamefont{Lezius}}, \bibinfo {author}
  {\bibfnamefont{K.~L.}\ \bibnamefont{Kompa}}, \bibinfo {author}
  {\bibfnamefont{H.}~\bibnamefont{Muller}}, \bibinfo {author}
  {\bibfnamefont{M.~J.~J.}\ \bibnamefont{Vrakking}}, \bibinfo {author}
  {\bibfnamefont{S.}~\bibnamefont{Hendel}}, \bibinfo {author}
  {\bibfnamefont{U.}~\bibnamefont{Kleineberg}}, \bibinfo {author}
  {\bibfnamefont{U.}~\bibnamefont{Heinzmann}}, \bibinfo {author}
  {\bibfnamefont{M.}~\bibnamefont{Drescher}},\ and\ \bibinfo {author}
  {\bibfnamefont{F.}~\bibnamefont{Krausz}},\ }%
  \bibfield{journal}{%
  \bibinfo {journal} {Nature}\ }%
  \textbf{\bibinfo {volume} {446}},\ \bibinfo {pages} {627} (\bibinfo {year}
  {2007})%
  \bibAnnoteFile{NoStop}{uiberacker_attosecond_2007}%
\bibitem{drescher_time-resolved_2002}%
  \BibitemOpen
  \bibfield{author}{%
  \bibinfo {author} {\bibfnamefont{M.}~\bibnamefont{Drescher}}, \bibinfo
  {author} {\bibfnamefont{M.}~\bibnamefont{Hentschel}}, \bibinfo {author}
  {\bibfnamefont{R.}~\bibnamefont{Kienberger}}, \bibinfo {author}
  {\bibfnamefont{M.}~\bibnamefont{Uiberacker}}, \bibinfo {author}
  {\bibfnamefont{V.}~\bibnamefont{Yakovlev}}, \bibinfo {author}
  {\bibfnamefont{A.}~\bibnamefont{Scrinzi}}, \bibinfo {author}
  {\bibfnamefont{T.}~\bibnamefont{Westerwalbesloh}}, \bibinfo {author}
  {\bibfnamefont{U.}~\bibnamefont{Kleineberg}}, \bibinfo {author}
  {\bibfnamefont{U.}~\bibnamefont{Heinzmann}},\ and\ \bibinfo {author}
  {\bibfnamefont{F.}~\bibnamefont{Krausz}},\ }%
  \bibfield{journal}{%
  \bibinfo {journal} {Nature}\ }%
  \textbf{\bibinfo {volume} {419}},\ \bibinfo {pages} {803} (\bibinfo {year}
  {2002})%
  \bibAnnoteFile{NoStop}{drescher_time-resolved_2002}%
\bibitem{schultze_delay_2010}%
  \BibitemOpen
  \bibfield{author}{%
  \bibinfo {author} {\bibfnamefont{M.}~\bibnamefont{Schultze}}, \bibinfo
  {author} {\bibfnamefont{M.}~\bibnamefont{Fie\ss{}}}, \bibinfo {author}
  {\bibfnamefont{N.}~\bibnamefont{Karpowicz}}, \bibinfo {author}
  {\bibfnamefont{J.}~\bibnamefont{Gagnon}}, \bibinfo {author}
  {\bibfnamefont{M.}~\bibnamefont{Korbman}}, \bibinfo {author}
  {\bibfnamefont{M.}~\bibnamefont{Hofstetter}}, \bibinfo {author}
  {\bibfnamefont{S.}~\bibnamefont{Neppl}}, \bibinfo {author}
  {\bibfnamefont{A.~L.}\ \bibnamefont{Cavalieri}}, \bibinfo {author}
  {\bibfnamefont{Y.}~\bibnamefont{Komninos}}, \bibinfo {author}
  {\bibfnamefont{T.}~\bibnamefont{Mercouris}}, \bibinfo {author}
  {\bibfnamefont{C.~A.}\ \bibnamefont{Nicolaides}}, \bibinfo {author}
  {\bibfnamefont{R.}~\bibnamefont{Pazourek}}, \bibinfo {author}
  {\bibfnamefont{S.}~\bibnamefont{Nagele}}, \bibinfo {author}
  {\bibfnamefont{J.}~\bibnamefont{Feist}}, \bibinfo {author}
  {\bibfnamefont{J.}~\bibnamefont{Burgd\"orfer}}, \bibinfo {author}
  {\bibfnamefont{A.~M.}\ \bibnamefont{Azzeer}}, \bibinfo {author}
  {\bibfnamefont{R.}~\bibnamefont{Ernstorfer}}, \bibinfo {author}
  {\bibfnamefont{R.}~\bibnamefont{Kienberger}}, \bibinfo {author}
  {\bibfnamefont{U.}~\bibnamefont{Kleineberg}}, \bibinfo {author}
  {\bibfnamefont{E.}~\bibnamefont{Goulielmakis}}, \bibinfo {author}
  {\bibfnamefont{F.}~\bibnamefont{Krausz}},\ and\ \bibinfo {author}
  {\bibfnamefont{V.~S.}\ \bibnamefont{Yakovlev}},\ }%
  \bibfield{journal}{%
  \bibinfo {journal} {Science}\ }%
  \textbf{\bibinfo {volume} {328}},\ \bibinfo {pages} {1658} (\bibinfo {year}
  {2010})%
  \bibAnnoteFile{NoStop}{schultze_delay_2010}%
\bibitem{wheatstone1834}%
  \BibitemOpen
  \bibfield{author}{%
  \bibinfo {author} {\bibfnamefont{C.}~\bibnamefont{Wheatstone}},\ }%
  \bibfield{journal}{%
  \bibinfo {journal} {Phil. Tans. R. Soc. Lond.}\ }%
  \textbf{\bibinfo {volume} {124}},\ \bibinfo {pages} {583} (\bibinfo {year}
  {1834})%
  \bibAnnoteFile{NoStop}{wheatstone1834}%
\bibitem{bradley1971}%
  \BibitemOpen
  \bibfield{author}{%
  \bibinfo {author} {\bibfnamefont{D.~J.}\ \bibnamefont{Bradley}}, \bibinfo
  {author} {\bibfnamefont{B.}~\bibnamefont{Liddy}},\ and\ \bibinfo {author}
  {\bibfnamefont{W.}~\bibnamefont{Sleat}},\ }%
  \bibfield{journal}{%
  \bibinfo {journal} {Opt. Comm.}\ }%
  \textbf{\bibinfo {volume} {2}},\ \bibinfo {pages} {391} (\bibinfo {year}
  {1971})%
  \bibAnnoteFile{NoStop}{bradley1971}%
\bibitem{feng_x-ray_2007}%
  \BibitemOpen
  \bibfield{author}{%
  \bibinfo {author} {\bibfnamefont{J.}~\bibnamefont{Feng}}, \bibinfo {author}
  {\bibfnamefont{H.~J.}\ \bibnamefont{Shin}}, \bibinfo {author}
  {\bibfnamefont{J.~R.}\ \bibnamefont{Nasiatka}}, \bibinfo {author}
  {\bibfnamefont{W.}~\bibnamefont{Wan}}, \bibinfo {author}
  {\bibfnamefont{A.~T.}\ \bibnamefont{Young}}, \bibinfo {author}
  {\bibfnamefont{G.}~\bibnamefont{Huang}}, \bibinfo {author}
  {\bibfnamefont{A.}~\bibnamefont{Comin}}, \bibinfo {author}
  {\bibfnamefont{J.}~\bibnamefont{Byrd}},\ and\ \bibinfo {author}
  {\bibfnamefont{H.~A.}\ \bibnamefont{Padmore}},\ }%
  \bibfield{journal}{%
  \bibinfo {journal} {Appl. Phys. Lett.}\ }%
  \textbf{\bibinfo {volume} {91}},\ \bibinfo {pages} {134102} (\bibinfo {year}
  {2007})%
  \bibAnnoteFile{NoStop}{feng_x-ray_2007}%
\bibitem{kienberger_atomic_2004}%
  \BibitemOpen
  \bibfield{author}{%
  \bibinfo {author} {\bibfnamefont{R.}~\bibnamefont{Kienberger}}, \bibinfo
  {author} {\bibfnamefont{E.}~\bibnamefont{Goulielmakis}}, \bibinfo {author}
  {\bibfnamefont{M.}~\bibnamefont{Uiberacker}}, \bibinfo {author}
  {\bibfnamefont{A.}~\bibnamefont{Baltuska}}, \bibinfo {author}
  {\bibfnamefont{V.}~\bibnamefont{Yakovlev}}, \bibinfo {author}
  {\bibfnamefont{F.}~\bibnamefont{Bammer}}, \bibinfo {author}
  {\bibfnamefont{A.}~\bibnamefont{Scrinzi}}, \bibinfo {author}
  {\bibfnamefont{T.}~\bibnamefont{Westerwalbesloh}}, \bibinfo {author}
  {\bibfnamefont{U.}~\bibnamefont{Kleineberg}}, \bibinfo {author}
  {\bibfnamefont{U.}~\bibnamefont{Heinzmann}}, \bibinfo {author}
  {\bibfnamefont{M.}~\bibnamefont{Drescher}},\ and\ \bibinfo {author}
  {\bibfnamefont{F.}~\bibnamefont{Krausz}},\ }%
  \bibfield{journal}{%
  \bibinfo {journal} {Nature}\ }%
  \textbf{\bibinfo {volume} {427}},\ \bibinfo {pages} {817} (\bibinfo {year}
  {2004})%
  \bibAnnoteFile{NoStop}{kienberger_atomic_2004}%
\bibitem{itatani_attosecond_2002}%
  \BibitemOpen
  \bibfield{author}{%
  \bibinfo {author} {\bibfnamefont{J.}~\bibnamefont{Itatani}}, \bibinfo
  {author} {\bibfnamefont{F.}~\bibnamefont{Qu\'{e}r\'{e}}}, \bibinfo {author}
  {\bibfnamefont{G.~L.}\ \bibnamefont{Yudin}}, \bibinfo {author}
  {\bibfnamefont{M.~Y.}\ \bibnamefont{Ivanov}}, \bibinfo {author}
  {\bibfnamefont{F.}~\bibnamefont{Krausz}},\ and\ \bibinfo {author}
  {\bibfnamefont{P.~B.}\ \bibnamefont{Corkum}},\ }%
  \bibfield{journal}{%
  \bibinfo {journal} {Phys. Rev. Lett.}\ }%
  \textbf{\bibinfo {volume} {88}},\ \bibinfo {pages} {173903} (\bibinfo {year}
  {2002})%
  \bibAnnoteFile{NoStop}{itatani_attosecond_2002}%
\bibitem{ipp_streaking_2011}%
  \BibitemOpen
  \bibfield{author}{%
  \bibinfo {author} {\bibfnamefont{A.}~\bibnamefont{Ipp}}, \bibinfo {author}
  {\bibfnamefont{J.}~\bibnamefont{Evers}}, \bibinfo {author}
  {\bibfnamefont{C.~H.}\ \bibnamefont{Keitel}},\ and\ \bibinfo {author}
  {\bibfnamefont{K.~Z.}\ \bibnamefont{Hatsagortsyan}},\ }%
  \bibfield{journal}{%
  \bibinfo {journal} {Physics Letters B}\ }%
  \textbf{\bibinfo {volume} {702}},\ \bibinfo {pages} {383} (\bibinfo {year}
  {2011})%
  \bibAnnoteFile{NoStop}{ipp_streaking_2011}%
\bibitem{fruhling_single-shot_2009}%
  \BibitemOpen
  \bibfield{author}{%
  \bibinfo {author} {\bibfnamefont{U.}~\bibnamefont{Fr\"uhling}}, \bibinfo
  {author} {\bibfnamefont{M.}~\bibnamefont{Wieland}}, \bibinfo {author}
  {\bibfnamefont{M.}~\bibnamefont{Gensch}}, \bibinfo {author}
  {\bibfnamefont{T.}~\bibnamefont{Gebert}}, \bibinfo {author}
  {\bibfnamefont{B.}~\bibnamefont{Sch\"utte}}, \bibinfo {author}
  {\bibfnamefont{M.}~\bibnamefont{Krikunova}}, \bibinfo {author}
  {\bibfnamefont{R.}~\bibnamefont{Kalms}}, \bibinfo {author}
  {\bibfnamefont{F.}~\bibnamefont{Budzyn}}, \bibinfo {author}
  {\bibfnamefont{O.}~\bibnamefont{Grimm}}, \bibinfo {author}
  {\bibfnamefont{J.}~\bibnamefont{Rossbach}}, \bibinfo {author}
  {\bibfnamefont{E.}~\bibnamefont{Pl\"onjes}},\ and\ \bibinfo {author}
  {\bibfnamefont{M.}~\bibnamefont{Drescher}},\ }%
  \bibfield{journal}{%
  \bibinfo {journal} {Nat. Photon.}\ }%
  \textbf{\bibinfo {volume} {3}},\ \bibinfo {pages} {523} (\bibinfo {year}
  {2009})%
  \bibAnnoteFile{NoStop}{fruhling_single-shot_2009}%
\bibitem{krasovskii_spectral_2007}%
  \BibitemOpen
  \bibfield{author}{%
  \bibinfo {author} {\bibfnamefont{E.~E.}\ \bibnamefont{Krasovskii}}\ and\
  \bibinfo {author} {\bibfnamefont{M.}~\bibnamefont{Bonitz}},\ }%
  \bibfield{journal}{%
  \bibinfo {journal} {Phys. Rev. Lett.}\ }%
  \textbf{\bibinfo {volume} {99}},\ \bibinfo {pages} {247601} (\bibinfo {year}
  {2007})%
  \bibAnnoteFile{NoStop}{krasovskii_spectral_2007}%
\bibitem{krasovskii_towards_2009}%
  \BibitemOpen
  \bibfield{author}{%
  \bibinfo {author} {\bibfnamefont{E.~E.}\ \bibnamefont{Krasovskii}}\ and\
  \bibinfo {author} {\bibfnamefont{M.}~\bibnamefont{Bonitz}},\ }%
  \bibfield{journal}{%
  \bibinfo {journal} {Phys. Rev. A}\ }%
  \textbf{\bibinfo {volume} {80}},\ \bibinfo {pages} {053421} (\bibinfo {year}
  {2009})%
  \bibAnnoteFile{NoStop}{krasovskii_towards_2009}%
\bibitem{meitner_ueber_1922}%
  \BibitemOpen
  \bibfield{author}{%
  \bibinfo {author} {\bibfnamefont{L.}~\bibnamefont{Meitner}},\ }%
  \bibfield{journal}{%
  \bibinfo {journal} {Zeitschrift f\"ur Physik}\ }%
  \textbf{\bibinfo {volume} {11}},\ \bibinfo {pages} {35} (\bibinfo {year}
  {1922})%
  \bibAnnoteFile{NoStop}{meitner_ueber_1922}%
\bibitem{auger_sur_1925}%
  \BibitemOpen
  \bibfield{author}{%
  \bibinfo {author} {\bibfnamefont{P.}~\bibnamefont{Auger}},\ }%
  \bibfield{journal}{%
  \bibinfo {journal} {Comptes Rendus}\ }%
  \textbf{\bibinfo {volume} {180}},\ \bibinfo {pages} {65} (\bibinfo {year}
  {1925})%
  \bibAnnoteFile{NoStop}{auger_sur_1925}%
\bibitem{schuette_evidence_2011}%
  \BibitemOpen
  \bibfield{author}{%
  \bibinfo {author} {\bibfnamefont{B.}~\bibnamefont{Sch\"utte}}, \bibinfo
  {author} {\bibfnamefont{S.}~\bibnamefont{Bauch}}, \bibinfo {author}
  {\bibfnamefont{U.}~\bibnamefont{Fr\"uhling}}, \bibinfo {author}
  {\bibfnamefont{M.}~\bibnamefont{Wieland}}, \bibinfo {author}
  {\bibfnamefont{M.}~\bibnamefont{Gensch}}, \bibinfo {author}
  {\bibfnamefont{E.}~\bibnamefont{Pl\"onjes}}, \bibinfo {author}
  {\bibfnamefont{T.}~\bibnamefont{Gaumnitz}}, \bibinfo {author}
  {\bibfnamefont{A.}~\bibnamefont{Azima}}, \bibinfo {author}
  {\bibfnamefont{M.}~\bibnamefont{Bonitz}},\ and\ \bibinfo {author}
  {\bibfnamefont{M.}~\bibnamefont{Drescher}},\ }%
  \bibinfo {journal} {accepted for publication in Phys. Rev. Lett. 2012}%
  \bibAnnoteFile{NoStop}{schuette_evidence_2011}%
\bibitem{schuette_electron_2011}%
  \BibitemOpen
\bibfield{journal}{%
    }%
  \bibfield{author}{%
  \bibinfo {author} {\bibfnamefont{B.}~\bibnamefont{Sch\"utte}}, \bibinfo
  {author} {\bibfnamefont{U.}~\bibnamefont{Fr\"uhling}}, \bibinfo {author}
  {\bibfnamefont{M.}~\bibnamefont{Wieland}}, \bibinfo {author}
  {\bibfnamefont{A.}~\bibnamefont{Azima}},\ and\ \bibinfo {author}
  {\bibfnamefont{M.}~\bibnamefont{Drescher}},\ }%
  \bibfield{journal}{%
  \bibinfo {journal} {Opt. Express}\ }%
  \textbf{\bibinfo {volume} {19}},\ \bibinfo {pages} {18833} (\bibinfo {year}
  {2011})%
  \bibAnnoteFile{NoStop}{schuette_electron_2011}%
\bibitem{niehaus_analysis_1977}%
  \BibitemOpen
  \bibfield{author}{%
  \bibinfo {author} {\bibfnamefont{A.}~\bibnamefont{Niehaus}},\ }%
  \bibfield{journal}{%
  \bibinfo {journal} {J. Phys. B: At. Mol. Phys.}\ }%
  \textbf{\bibinfo {volume} {10}},\ \bibinfo {pages} {1845} (\bibinfo {year}
  {1977})%
  \bibAnnoteFile{NoStop}{niehaus_analysis_1977}%
\bibitem{ogurtsov_auger_1983}%
  \BibitemOpen
  \bibfield{author}{%
  \bibinfo {author} {\bibfnamefont{G.~N.}\ \bibnamefont{Ogurtsov}},\ }%
  \bibfield{journal}{%
  \bibinfo {journal} {J. Phys. B: At. Mol. Phys.}\ }%
  \textbf{\bibinfo {volume} {16}},\ \bibinfo {pages} {L745} (\bibinfo {year}
  {1983})%
  \bibAnnoteFile{NoStop}{ogurtsov_auger_1983}%
\bibitem{russek_post-collision_1986}%
  \BibitemOpen
  \bibfield{author}{%
  \bibinfo {author} {\bibfnamefont{A.}~\bibnamefont{Russek}}\ and\ \bibinfo
  {author} {\bibfnamefont{W.}~\bibnamefont{Mehlhorn}},\ }%
  \bibfield{journal}{%
  \bibinfo {journal} {J. Phys. B: At. Mol. Phys.}\ }%
  \textbf{\bibinfo {volume} {19}},\ \bibinfo {pages} {911} (\bibinfo {year}
  {1986})%
  \bibAnnoteFile{NoStop}{russek_post-collision_1986}%
\bibitem{kuchiev_resonant_1986}%
  \BibitemOpen
  \bibfield{author}{%
  \bibinfo {author} {\bibfnamefont{M.~Y.}\ \bibnamefont{Kuchiev}}\ and\
  \bibinfo {author} {\bibfnamefont{S.}~\bibnamefont{Sheinerman}},\ }%
  \bibfield{journal}{%
  \bibinfo {journal} {Sov. Phys. JETP}\ }%
  \textbf{\bibinfo {volume} {63}},\ \bibinfo {pages} {986} (\bibinfo {year}
  {1986})%
  \bibAnnoteFile{NoStop}{kuchiev_resonant_1986}%
\bibitem{kuchiev_post-collision_1989}%
  \BibitemOpen
  \bibfield{author}{%
  \bibinfo {author} {\bibfnamefont{M.~Y.}\ \bibnamefont{Kuchiev}}\ and\
  \bibinfo {author} {\bibfnamefont{S.}~\bibnamefont{Sheinerman}},\ }%
  \bibfield{journal}{%
  \bibinfo {journal} {Sov. Phys. Uspekhi}\ }%
  \textbf{\bibinfo {volume} {158}},\ \bibinfo {pages} {353} (\bibinfo {year}
  {1989})%
  \bibAnnoteFile{NoStop}{kuchiev_post-collision_1989}%
\bibitem{aberg_unified_1992}%
  \BibitemOpen
  \bibfield{author}{%
  \bibinfo {author} {\bibfnamefont{T.}~\bibnamefont{\r{A}berg}},\ }%
  \bibfield{journal}{%
  \bibinfo {journal} {Physica Scripta}\ }%
  \textbf{\bibinfo {volume} {T41}},\ \bibinfo {pages} {71} (\bibinfo {year}
  {1992})%
  \bibAnnoteFile{NoStop}{aberg_unified_1992}%
\bibitem{haan_numerical_1994}%
  \BibitemOpen
  \bibfield{author}{%
  \bibinfo {author} {\bibfnamefont{S.~L.}\ \bibnamefont{Haan}}, \bibinfo
  {author} {\bibfnamefont{R.}~\bibnamefont{Grobe}},\ and\ \bibinfo {author}
  {\bibfnamefont{J.~H.}\ \bibnamefont{Eberly}},\ }%
  \bibfield{journal}{%
  \bibinfo {journal} {Phys. Rev. A}\ }%
  \textbf{\bibinfo {volume} {50}},\ \bibinfo {pages} {378} (\bibinfo {year}
  {1994})%
  \bibAnnoteFile{NoStop}{haan_numerical_1994}%
\bibitem{hu_time-dependent_2005}%
  \BibitemOpen
  \bibfield{author}{%
  \bibinfo {author} {\bibfnamefont{S.~X.}\ \bibnamefont{Hu}}\ and\ \bibinfo
  {author} {\bibfnamefont{L.~A.}\ \bibnamefont{Collins}},\ }%
  \bibfield{journal}{%
  \bibinfo {journal} {Phys. Rev. A}\ }%
  \textbf{\bibinfo {volume} {71}},\ \bibinfo {pages} {062707} (\bibinfo {year}
  {2005})%
  \bibAnnoteFile{NoStop}{hu_time-dependent_2005}%
\bibitem{fano_effects_1961}%
  \BibitemOpen
  \bibfield{author}{%
  \bibinfo {author} {\bibfnamefont{U.}~\bibnamefont{Fano}},\ }%
  \bibfield{journal}{%
  \bibinfo {journal} {Phys Rev.}\ }%
  \textbf{\bibinfo {volume} {124}},\ \bibinfo {pages} {1866} (\bibinfo {year}
  {1961})%
  \bibAnnoteFile{NoStop}{fano_effects_1961}%
\bibitem{kazansky_nonstationary_2005}%
  \BibitemOpen
  \bibfield{author}{%
  \bibinfo {author} {\bibfnamefont{A.~K.}\ \bibnamefont{Kazansky}}\ and\
  \bibinfo {author} {\bibfnamefont{N.~M.}\ \bibnamefont{Kabachnik}},\ }%
  \bibfield{journal}{%
  \bibinfo {journal} {Phy. Rev. A}\ }%
  \textbf{\bibinfo {volume} {72}},\ \bibinfo {pages} {052714} (\bibinfo {year}
  {2005})%
  \bibAnnoteFile{NoStop}{kazansky_nonstationary_2005}%
\bibitem{kazansky_triple_2006}%
  \BibitemOpen
  \bibfield{author}{%
  \bibinfo {author} {\bibfnamefont{A.~K.}\ \bibnamefont{Kazansky}}\ and\
  \bibinfo {author} {\bibfnamefont{N.~M.}\ \bibnamefont{Kabachnik}},\ }%
  \bibfield{journal}{%
  \bibinfo {journal} {Phys. Rev. A}\ }%
  \textbf{\bibinfo {volume} {73}},\ \bibinfo {pages} {062712} (\bibinfo {year}
  {2006})%
  \bibAnnoteFile{NoStop}{kazansky_triple_2006}%
\bibitem{kazansky_time-dependent_2009}%
  \BibitemOpen
  \bibfield{author}{%
  \bibinfo {author} {\bibfnamefont{A.~K.}\ \bibnamefont{Kazansky}}, \bibinfo
  {author} {\bibfnamefont{I.~P.}\ \bibnamefont{Sazhina}},\ and\ \bibinfo
  {author} {\bibfnamefont{N.~M.}\ \bibnamefont{Kabachnik}},\ }%
  \bibfield{journal}{%
  \bibinfo {journal} {J. Phys. B: At. Mol. Opt. Phys.}\ }%
  \textbf{\bibinfo {volume} {42}},\ \bibinfo {pages} {245601} (\bibinfo {year}
  {2009})%
  \bibAnnoteFile{NoStop}{kazansky_time-dependent_2009}%
\bibitem{buth_theory_2009}%
  \BibitemOpen
  \bibfield{author}{%
  \bibinfo {author} {\bibfnamefont{C.}~\bibnamefont{Buth}}\ and\ \bibinfo
  {author} {\bibfnamefont{K.}~\bibnamefont{Schafer}},\ }%
  \bibfield{journal}{%
  \bibinfo {journal} {Phys. Rev. A}\ }%
  \textbf{\bibinfo {volume} {80}},\ \bibinfo {pages} {033410} (\bibinfo {year}
  {2009})%
  \bibAnnoteFile{NoStop}{buth_theory_2009}%
\bibitem{wickenhauser_theoretical_2004}%
  \BibitemOpen
  \bibfield{author}{%
  \bibinfo {author} {\bibfnamefont{M.}~\bibnamefont{Wickenhauser}}\ and\
  \bibinfo {author} {\bibfnamefont{J.}~\bibnamefont{Burgd\"{o}rfer}},\ }%
  \bibfield{journal}{%
  \bibinfo {journal} {Laser Physics}\ }%
  \textbf{\bibinfo {volume} {14}},\ \bibinfo {pages} {492} (\bibinfo {year}
  {2004})%
  \bibAnnoteFile{NoStop}{wickenhauser_theoretical_2004}%
\bibitem{wickenhauser_time_2005}%
  \BibitemOpen
  \bibfield{author}{%
  \bibinfo {author} {\bibfnamefont{M.}~\bibnamefont{Wickenhauser}}, \bibinfo
  {author} {\bibfnamefont{J.}~\bibnamefont{Burgd\"{o}rfer}}, \bibinfo {author}
  {\bibfnamefont{F.}~\bibnamefont{Krausz}},\ and\ \bibinfo {author}
  {\bibfnamefont{M.}~\bibnamefont{Drescher}},\ }%
  \bibfield{journal}{%
  \bibinfo {journal} {Phys. Rev. Lett.}\ }%
  \textbf{\bibinfo {volume} {94}},\ \bibinfo {pages} {023002} (\bibinfo {year}
  {2005})%
  \bibAnnoteFile{NoStop}{wickenhauser_time_2005}%
\bibitem{hergenhahn_population_2006}%
  \BibitemOpen
  \bibfield{author}{%
  \bibinfo {author} {\bibfnamefont{U.}~\bibnamefont{Hergenhahn}}, \bibinfo
  {author} {\bibfnamefont{A.}~\bibnamefont{{D}e Fanis}}, \bibinfo {author}
  {\bibfnamefont{G.}~\bibnamefont{Pr\"{u}mper}}, \bibinfo {author}
  {\bibfnamefont{A.~K.}\ \bibnamefont{Kazansky}}, \bibinfo {author}
  {\bibfnamefont{N.~M.}\ \bibnamefont{Kabachnik}},\ and\ \bibinfo {author}
  {\bibfnamefont{K.}~\bibnamefont{Ueda}},\ }%
  \bibfield{journal}{%
  \bibinfo {journal} {Phys. Rev. A}\ }%
  \textbf{\bibinfo {volume} {73}},\ \bibinfo {pages} {022709} (\bibinfo {year}
  {2006})%
  \bibAnnoteFile{NoStop}{hergenhahn_population_2006}%
\bibitem{hergenhahn_study_2005}%
  \BibitemOpen
  \bibfield{author}{%
  \bibinfo {author} {\bibfnamefont{U.}~\bibnamefont{Hergenhahn}}, \bibinfo
  {author} {\bibfnamefont{A.}~\bibnamefont{{D}e Fanis}}, \bibinfo {author}
  {\bibfnamefont{G.}~\bibnamefont{Pr\"{u}mper}}, \bibinfo {author}
  {\bibfnamefont{A.~K.}\ \bibnamefont{Kazansky}}, \bibinfo {author}
  {\bibfnamefont{N.~M.}\ \bibnamefont{Kabachnik}},\ and\ \bibinfo {author}
  {\bibfnamefont{K.}~\bibnamefont{Ueda}},\ }%
  \bibfield{journal}{%
  \bibinfo {journal} {J. Phys. B: At. Mol. Opt. Phys.}\ }%
  \textbf{\bibinfo {volume} {38}},\ \bibinfo {pages} {2843} (\bibinfo {year}
  {2005})%
  \bibAnnoteFile{NoStop}{hergenhahn_study_2005}%
\bibitem{kazansky_sideband_2009}%
  \BibitemOpen
  \bibfield{author}{%
  \bibinfo {author} {\bibfnamefont{A.~K.}\ \bibnamefont{Kazansky}}\ and\
  \bibinfo {author} {\bibfnamefont{N.~M.}\ \bibnamefont{Kabachnik}},\ }%
  \bibfield{journal}{%
  \bibinfo {journal} {J. Phys. B: At. Mol. Opt. Phys.}\ }%
  \textbf{\bibinfo {volume} {42}},\ \bibinfo {pages} {121002} (\bibinfo {year}
  {2009})%
  \bibAnnoteFile{NoStop}{kazansky_sideband_2009}%
\bibitem{kazansky_gross_2010}%
  \BibitemOpen
  \bibfield{author}{%
  \bibinfo {author} {\bibfnamefont{A.~K.}\ \bibnamefont{Kazansky}}\ and\
  \bibinfo {author} {\bibfnamefont{N.~M.}\ \bibnamefont{Kabachnik}},\ }%
  \bibfield{journal}{%
  \bibinfo {journal} {J. Phys. B: At. Mol. Opt. Phys.}\ }%
  \textbf{\bibinfo {volume} {43}},\ \bibinfo {pages} {035601} (\bibinfo {year}
  {2010})%
  \bibAnnoteFile{NoStop}{kazansky_gross_2010}%
\bibitem{su_model_1991}%
  \BibitemOpen
  \bibfield{author}{%
  \bibinfo {author} {\bibfnamefont{Q.}~\bibnamefont{Su}}\ and\ \bibinfo
  {author} {\bibfnamefont{J.}~\bibnamefont{Eberly}},\ }%
  \bibfield{journal}{%
  \bibinfo {journal} {Phys. Rev. A}\ }%
  \textbf{\bibinfo {volume} {44}},\ \bibinfo {pages} {5997} (\bibinfo {year}
  {1991})%
  \bibAnnoteFile{NoStop}{su_model_1991}%
\bibitem{bauch_electronic_2010}%
  \BibitemOpen
  \bibfield{author}{%
  \bibinfo {author} {\bibfnamefont{S.}~\bibnamefont{Bauch}}, \bibinfo {author}
  {\bibfnamefont{K.}~\bibnamefont{Balzer}},\ and\ \bibinfo {author}
  {\bibfnamefont{M.}~\bibnamefont{Bonitz}},\ }%
  \bibfield{journal}{%
  \bibinfo {journal} {Europhys. Lett.}\ }%
  \textbf{\bibinfo {volume} {91}},\ \bibinfo {pages} {53001} (\bibinfo {year}
  {2010})%
  \bibAnnoteFile{NoStop}{bauch_electronic_2010}%
\bibitem{rescigno_numerical_2000}%
  \BibitemOpen
  \bibfield{author}{%
  \bibinfo {author} {\bibfnamefont{T.~N.}\ \bibnamefont{Rescigno}}\ and\
  \bibinfo {author} {\bibfnamefont{C.~W.}\ \bibnamefont{McCurdy}},\ }%
  \bibfield{journal}{%
  \bibinfo {journal} {Phys. Rev. A}\ }%
  \textbf{\bibinfo {volume} {62}},\ \bibinfo {pages} {032706} (\bibinfo {year}
  {2000})%
  \bibAnnoteFile{NoStop}{rescigno_numerical_2000}%
\bibitem{schneider_parallel_2006}%
  \BibitemOpen
  \bibfield{author}{%
  \bibinfo {author} {\bibfnamefont{B.~I.}\ \bibnamefont{Schneider}}, \bibinfo
  {author} {\bibfnamefont{L.~A.}\ \bibnamefont{Collins}},\ and\ \bibinfo
  {author} {\bibfnamefont{S.~X.}\ \bibnamefont{Hu}},\ }%
  \bibfield{journal}{%
  \bibinfo {journal} {Phys. Rev. E}\ }%
  \textbf{\bibinfo {volume} {73}},\ \bibinfo {pages} {036708} (\bibinfo {year}
  {2006})%
  \bibAnnoteFile{NoStop}{schneider_parallel_2006}%
\bibitem{snell_angular_2000}%
  \BibitemOpen
  \bibfield{author}{%
  \bibinfo {author} {\bibfnamefont{G.}~\bibnamefont{Snell}}, \bibinfo {author}
  {\bibfnamefont{E.}~\bibnamefont{Kukk}}, \bibinfo {author}
  {\bibfnamefont{B.}~\bibnamefont{Langer}},\ and\ \bibinfo {author}
  {\bibfnamefont{N.}~\bibnamefont{Berrah}},\ }%
  \bibfield{journal}{%
  \bibinfo {journal} {Phys. Rev. A}\ }%
  \textbf{\bibinfo {volume} {61}},\ \bibinfo {pages} {042709} (\bibinfo {year}
  {2000})%
  \bibAnnoteFile{NoStop}{snell_angular_2000}%
\bibitem{wolfram_sphere}%
  \BibitemOpen
  \bibfield{author}{%
  \bibinfo {author} {\bibfnamefont{E.~W.}\ \bibnamefont{Weisstein}},\ }%
  \bibfield{journal}{%
  \bibinfo {journal} {From MathWorld - A Wolfram Web Resource}}%
   (\bibinfo {year} {2011}),\
  \url{http://mathworld.wolfram.com/SpherePointPicking.html}%
  \bibAnnoteFile{NoStop}{wolfram_sphere}%
\bibitem{ott_md_2010}%
  \BibitemOpen
  \bibfield{author}{%
  \bibinfo {author} {\bibfnamefont{T.}~\bibnamefont{Ott}}, \bibinfo {author}
  {\bibfnamefont{P.}~\bibnamefont{Ludwig}}, \bibinfo {author}
  {\bibfnamefont{H.}~\bibnamefont{K\"ahlert}},\ and\ \bibinfo {author}
  {\bibfnamefont{M.}~\bibnamefont{Bonitz}},\ }%
  \bibfield{journal}{%
  \bibinfo {journal} {in M. Bonitz, N. Horing and P. Ludwig (eds.), Springer
  Series Atomic, Optical and Plasma Physics}\ }%
  \textbf{\bibinfo {volume} {59}} (\bibinfo {year} {2010})%
  \bibAnnoteFile{NoStop}{ott_md_2010}%
\end{thebibliography}

%

\end {document}